\documentclass[article]{IEEEtran}
\usepackage{booktabs}
\usepackage{cuted}
\usepackage{amsfonts}
\usepackage{amsmath}
\usepackage{eucal}
\usepackage{mathtools}
\usepackage{nccmath}
\usepackage{optidef}
\usepackage{tikz}
\usepackage{xfrac}
\usepackage{yhmath}
\usepackage{url}
\ifCLASSOPTIONcompsoc
    \usepackage[caption=false, font=normalsize, labelfont=sf, textfont=sf]{subfig}
\else
\usepackage[caption=false, font=footnotesize]{subfig}
\fi
\usepackage{multirow}

\usetikzlibrary{arrows}
\usetikzlibrary{decorations}
\usetikzlibrary{shapes}
\tikzset{
    block/.style   = {draw, thick, rectangle, minimum height = 2em},
    port/.style    = {draw, circle, minimum size=#1, inner sep=0pt, outer sep=0pt},
    root/.style    = {coordinate}
}

\usepackage{color,soul}

\definecolor{berkeleyblue}{rgb}{0,0.20,0.38}

\begin{document}

\title{Validation and Calibration of Energy Models with Real Vehicle Data from Chassis Dynamometer Experiments}


\author{
    \IEEEauthorblockN{Joy Carpio\IEEEauthorrefmark{1}\IEEEauthorrefmark{4}\thanks{Corresponding author. Email: engrjoycarpio@berkeley.edu}, 
                      Sulaiman Almatrudi\IEEEauthorrefmark{1}, 
                      Nour Khoudari\IEEEauthorrefmark{2}, 
                      Zhe Fu\IEEEauthorrefmark{1}, 
                      Kenneth Butts\IEEEauthorrefmark{3}, \\ 
                      Jonathan Lee\IEEEauthorrefmark{1}, 
                      Benjamin Seibold\IEEEauthorrefmark{2}, 
                      Alexandre Bayen\IEEEauthorrefmark{1}} \\

    \bigskip
    \IEEEauthorblockA{\IEEEauthorrefmark{1}University of California, Berkeley, USA}\\
    \IEEEauthorblockA{\IEEEauthorrefmark{2}Temple University, Philadelphia, Pennsylvania, USA}\\
    \IEEEauthorblockA{\IEEEauthorrefmark{3}Toyota North America}\\
    \IEEEauthorblockA{\IEEEauthorrefmark{4}National University, Manila, Philippines}
}

\maketitle

\begin{abstract}
Accurate estimation of vehicle fuel consumption typically requires detailed modeling of complex internal powertrain dynamics, often resulting in computationally intensive simulations. However, many transportation applications—such as traffic flow modeling, optimization, and control—require simplified models that are fast, interpretable, and easy to implement, while still maintaining fidelity to physical energy behavior. This work builds upon a recently developed model reduction pipeline that derives physics-like energy models from high-fidelity Autonomie vehicle simulations. These reduced models preserve essential vehicle dynamics, enabling realistic fuel consumption estimation with minimal computational overhead. While the reduced models have demonstrated strong agreement with their Autonomie counterparts, previous validation efforts have been confined to simulation environments. This study extends the validation by comparing the reduced energy model’s outputs against real-world vehicle data. Focusing on the MidSUV category, we tune the baseline Autonomie model to closely replicate the characteristics of a Toyota RAV4. We then assess the accuracy of the resulting reduced model in estimating fuel consumption under actual drive conditions. Our findings suggest that, when the reference Autonomie model is properly calibrated, the simplified model produced by the reduction pipeline can provide reliable, semi-principled fuel rate estimates suitable for large-scale transportation applications.
\end{abstract}

\IEEEpeerreviewmaketitle

\section{Introduction}
Numerous efforts have been made to demonstrate how controlled vehicles can stabilize traffic flow when mixed with human-driven cars. While many of these studies have relied on simulations~\cite{chou2022lord, delle2019feedback, bhadani2018dissipation, treiber2000congested, delle2019autonomous}, a few have attempted to validate this concept in real-world traffic scenarios~\cite{stern2018dissipation, sugiyama2008traffic, lee2021integrated}. One of the most recent and significant efforts to showcase the impact of controlled vehicles on the entire traffic flow is the CIRCLES project~\cite{lee2024traffic_control, CirclesSite, CIRCLES_AMR2020}.

A key objective of the Congestion Impact Reduction via CAV-in-the-loop Lagrangian Energy Smoothing (CIRCLES) project was to achieve a realistic quantification of energy consumption across various vehicle types, with a general goal of enhancing freeway traffic energy efficiency by at least 10\% \cite{CIRCLES_AMR2020, CirclesSite}. This ambitious initiative was conducted on the I20 freeway in Nashville, Tennessee, where a fleet of one hundred semi-autonomous vehicles, equipped with customized controllers and algorithms designed to mitigate stop-and-go traffic patterns, was deployed alongside human-driven vehicles~\cite{CirclesNewsArticle, lee2021integrated, lee2024traffic_control, lee2024traffic_smoothing}.

Research has shown that the default adaptive cruise control in vehicles does not achieve string stability on its own \cite{milanes2013cooperative, liang1999optimal, gunter2020commercially, shang2021impacts, xiao2010comprehensive}. To address this limitation, the CIRCLES project enhanced the default adaptive cruise control systems in the test vehicles by incorporating downstream traffic estimates. By leveraging various traffic mitigation strategies, including reinforcement learning, the project successfully aimed to reduce the traffic waves within the fleet.

One noteworthy outcome of the project is the development of mathematical models~\cite{khoudari2023reducing} that capture the physics-like properties of diverse representative vehicles. These models are expressed as simplified polynomial-fitted functions, taking into account instantaneous speed, acceleration, and road grade to predict fuel consumption in grams-per-second. The inherent simplicity of these energy models facilitates time-domain or moment-by-moment analysis, enabling a comprehensive examination of the energy consumption behavior for each vehicle type. Therefore, the evaluation of energy improvements achieved in the CIRCLES project's experiments heavily relies on the precision and reliability of these energy models. In addition to the CIRCLES project, these energy models can be highly effective in accurately assessing the energy impact of traffic mitigation strategies, such as those outlined in \cite{wu2024modifying, vinitsky2020energy, carpio2018traffic, lichtle2022deploying}, where direct access to the vehicle's trajectory is attained. This access can be achieved either through direct retrieval of CAN messages or by incorporating additional data processing layers, such as image processing, as demonstrated in \cite{GLOUDEMANS2023104311, carpio2018traffic, wu2019tracking}.

These energy models developed for the CIRCLES project underwent a meticulous derivation process employing detailed vehicle dynamics. These models were rigorously validated through Autonomie~\cite{autonomie}, a simulation tool crafted by Argonne National Laboratory, featuring comprehensive libraries delineating the intricate architectures of various vehicle types. To adapt these models for modeling, modifications were made to the Autonomie template Simulink-based models, enabling their execution in a \emph{Virtual Chassis Dynamometer (VCD)} -- a Matlab-created script emulating the testing process of an actual vehicle on a chassis dynamometer. This approach facilitated the identification of parametric properties specific to the chosen vehicle, which were then utilized to formulate mathematical models estimating fuel rate based on limited information: speed, acceleration, and road grade.

\begin{figure}
    \centering 
    \includegraphics[width=0.99\linewidth]{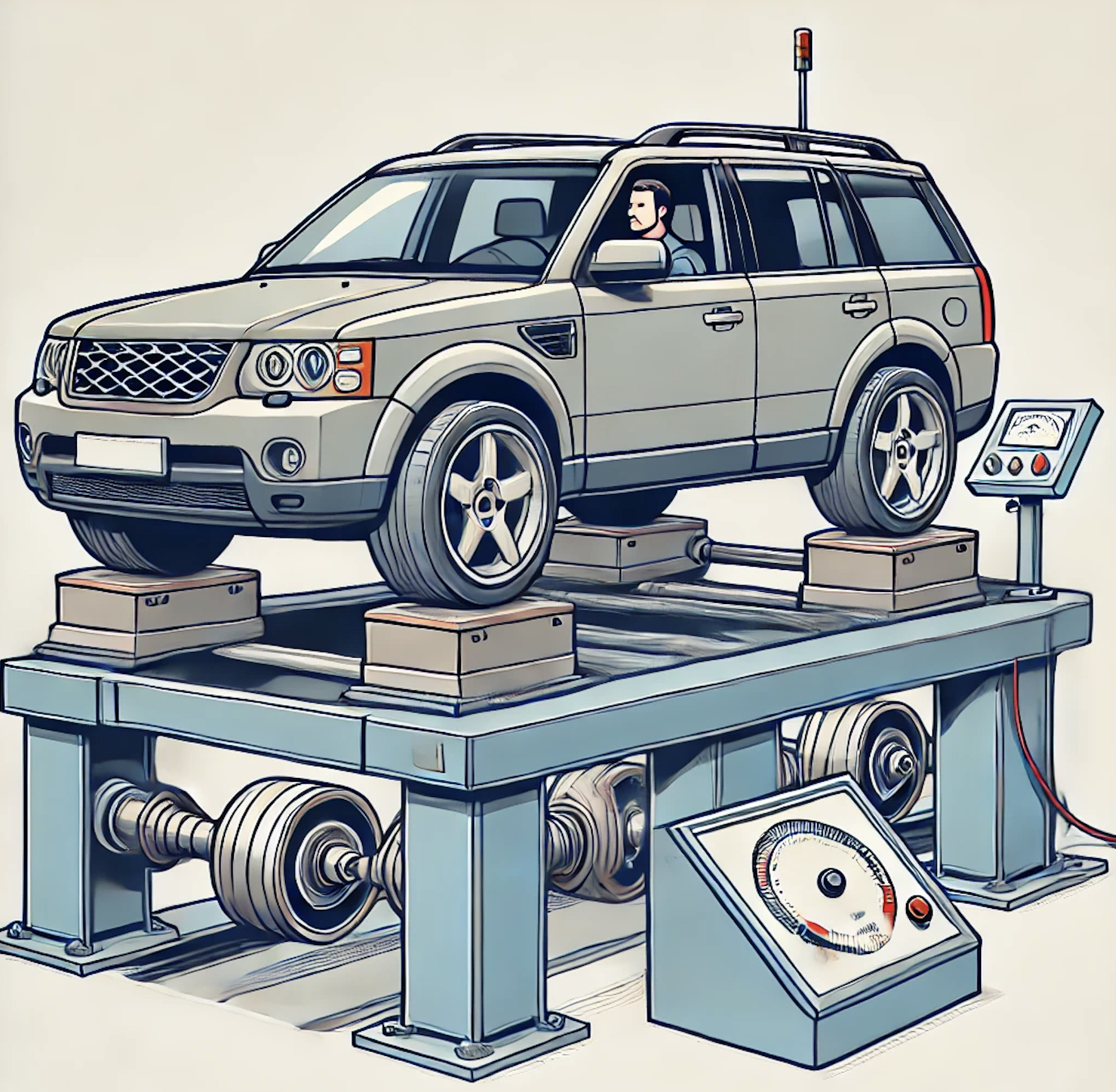}
    \caption{This image depicts a vehicle undergoing testing on a chassis dynamometer, with a human driver at the helm, executing the intended driving pattern \cite{Chassis_dyno}.}
    \label{Chassis_dyno}
\end{figure}

Nevertheless, in pursuit of enhanced validation and refinement, and to avoid restricting accuracy solely to Autonomie, an additional experiment was conducted. This experiment aimed to assess the precision of the energy models using real vehicle data. A tangible chassis dynamometer experiment was performed, employing a Toyota RAV4. An illustrative representation of a vehicle on a chassis dynamometer is presented in Figure \ref{Chassis_dyno}. This serves as a visual aid to explicate the experimental setup. Furthermore, for a more pertinent comparison, a specialized Autonomie model tailored to the Toyota RAV4 was derived from the Autonomie Midsize SUV template, the closest Autonomie model to the characteristic of Toyota RAV4. Consequently, the same mathematical models (semi-principled and simplifed models) were derived from this Toyota RAV4 Autonomie model, applying the modeling pipeline as described in \cite{khoudari2023reducing}.

The primary goal of the chassis dynamometer experiment was to validate the accuracy of the derived semi-principled and simplified energy models in comparison to actual vehicle data. This experiment also facilitated the fine-tuning process of the Autonomie model, consequently refining the semi-principled and simplified models, utilizing real-world vehicle performance.

\section{Problem Statement}
This study addresses the central problem of evaluating the fidelity of output produced by semi-principled and simplified energy models~\cite{khoudari2023reducing} in comparison to genuine vehicle data. A prior investigation successfully validated these energy models using Autonomie as a benchmark. In this paper, our primary goal is to scrutinize the performance of the energy models by contrasting their results with those derived from an operational vehicle subjected to identical input drive cycle conditions. Through this comparative analysis, the intention is to identify potential areas for improvement and refinement, ultimately aiming to enhance the precision of the models.

To achieve these objectives, it is essential to establish an experimental framework for collecting data from an actual vehicle. This setup must possess the adaptability to be utilized for both Autonomie and the energy models (both the semi-principled and simplified models), ensuring a consistent and standardized execution of the experiment across these platforms.

\section{The Energy Models: Autonomie, Semi-principled and Simplified Models}
\label{section:energy models}
\begin{figure}
    \centering 
    \includegraphics[width=0.99\linewidth]{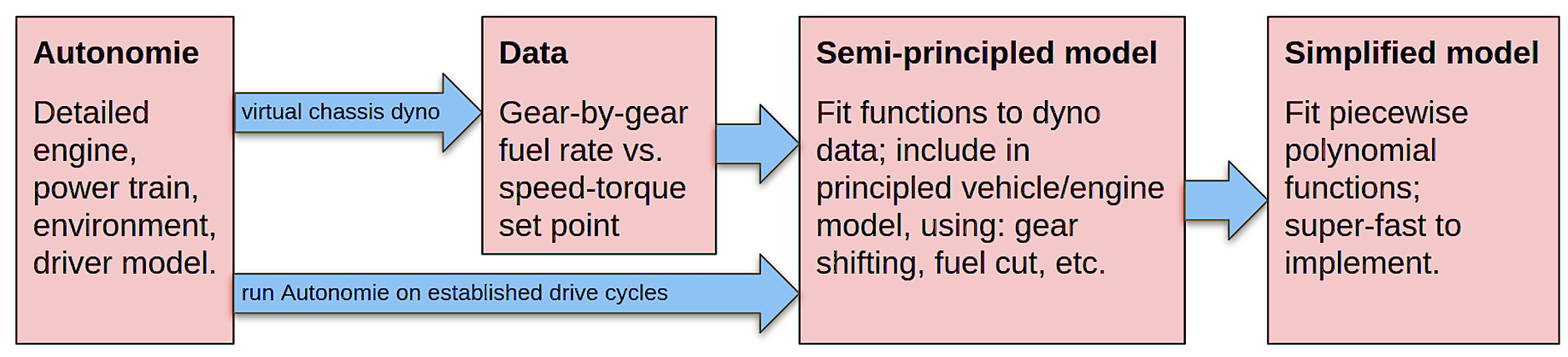}
    \caption{This diagram provides an overview of the energy modeling pipeline, beginning with Autonomie and culminating in the development of a simplified polynomial-fitted model.}
    \label{Process}
\end{figure}

This section describes the evolution of the energy models starting from utilizing a principled high-fidelity energy modeling tool, exemplified here by Autonomie. The baseline software serves as the foundation for generating data through the use of a virtual chassis dynamometer. Incorporating several vehicle parameters directly extracted from Autonomie -- such as mass, road load, and gear schedule -- fitted mappings between fuel rate and engine speed and torque are established. These mappings are then integrated into more straightforward and direct physical models, resulting in stateless physical models called the \emph{Semi-principled models}. Further simplification is applied to derive explicit formulas, eliminating the need for data tables, and producing the \emph{Simplified Models}. This simplification not only ensures rapid evaluation but also facilitates direct and convenient energy optimization. The resultant models harness the flexibility and accuracy of Autonomie, but with conceptual and structural simplicity, significantly reducing computational demands. This energy modeling pipeline is described in Figure \ref{Process}. A more detailed discussion of this pipeline is discussed in~\cite{khoudari2023reducing}.
\subsection{Autonomie Model}
\label{Autonomie Modeling}
The foundational source of the semi-principled and simplified models undergoing validation and refinement in this study is Autonomie. Autonomie possesses the capability to generate a fuel consumption report for a specific vehicle type under a given input driving pattern. The output takes the form of an instantaneous fuel rate at a specific timestamp. With confidence in the reliability of Autonomie, a semi-principled model was derived, grounded in the empirical concept that the instantaneous engine speed ($N$) and engine torque ($T$) can be correlated with fuel rate, while transmission output speed ($N_{\text{output}}$) and wheel force ($F_{\text{wheel}}$) can be linked to engine speed and torque. Both transmission output speed and wheel force are expressed as functions of speed, acceleration, and road grade.

\begin{figure}[htbp]
    \centering 
    \includegraphics[width=0.99\linewidth]{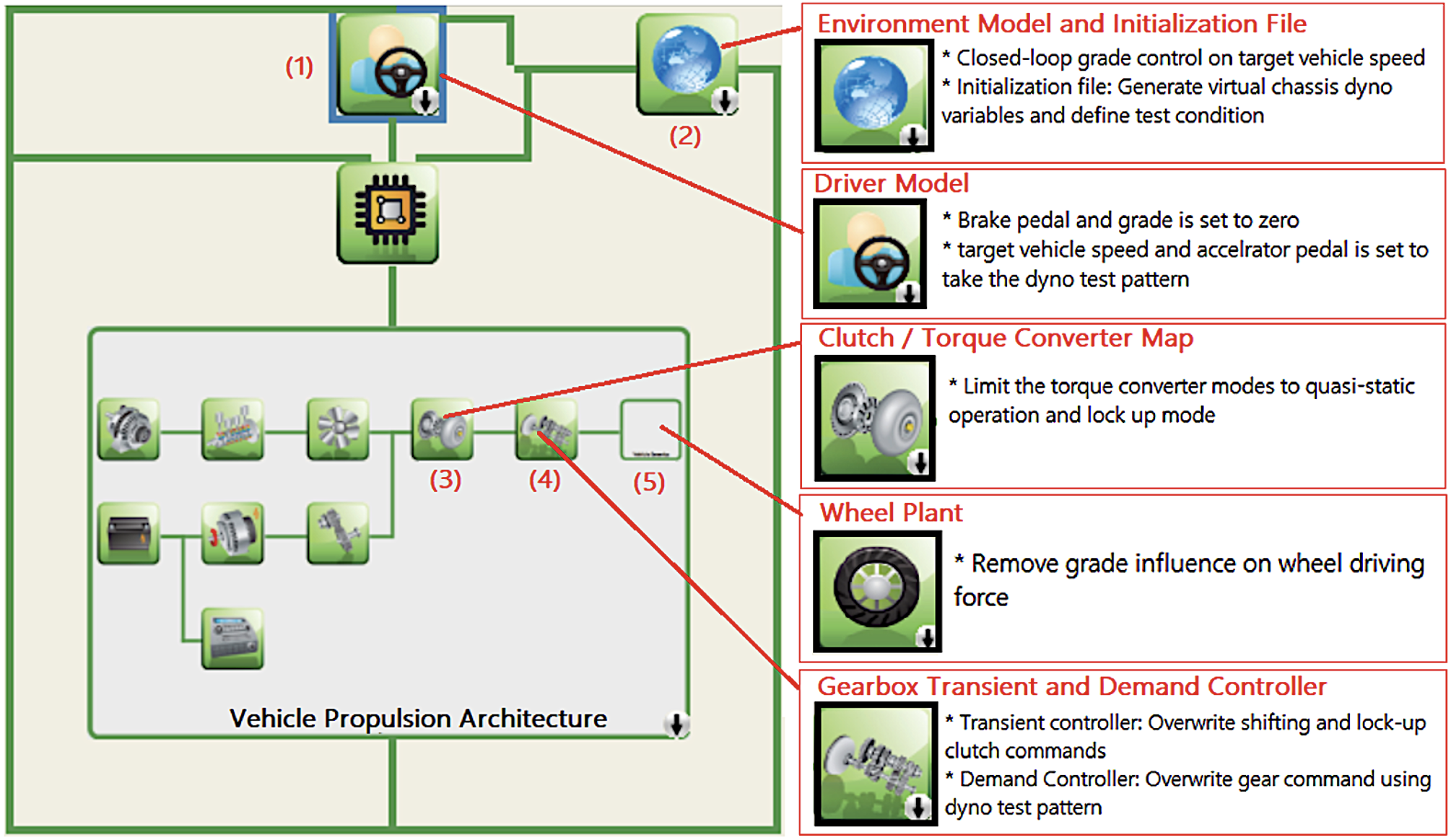}
    \caption{This figure represents the vehicle architecture in Autonomie, highlighting the specific components that require modification and detailing the necessary adaptations to extract the default vehicle parameters and maps essential for deriving the semi-principled model.}
    \label{Modification}
\end{figure}

To derive the essential maps, various components in the vehicle's template Autonomie Simulink models underwent customization, as illustrated in Figure \ref{Modification}. Several sections undergo modification to facilitate virtual chassis dynamometer operation. Firstly, in the \emph{Driver model}, the influence of the driver in applying brake and acceleration is eliminated by setting the brake pedal to zero and enforcing the accelerator pedal and target vehicle speed to follow the specified test pattern on the virtual chassis dynamometer. In the \emph{Environment model}, a closed-loop feedback controller is introduced to adjust the effective dynamometer force, ensuring that changes in road grade counteract wheel force variations induced by prescribed accelerator pedal changes. The initialization file incorporates virtual chassis dynamometer variables and relevant parameters extracted from Autonomie, along with the establishment of dyno test conditions. The \emph{Wheel plant} section is modified to cancel additional load produced by road grade, achieved by bypassing the block in the simulink model that takes the road grade as an input. The \emph{Gearbox controller} is adjusted to "steady gear" and "no gear shifting" conditions, maintaining normal engine mode determination, with target gear and torque converter bypass clutch schedules set in the environment model initialization file. Lastly, the \emph{Clutch/Torque converter} section restricts "Idle" and "Transient" modes in the torque converter, ensuring operation only in either transient acceleration or quasi-static modes. In a broader context, the primary objective of customizing the Simulink models is to capture maps intended for the semi-principled modeling process while ensuring that the captured behavior solely reflects the vehicle's energy consumption behavior.
\subsection{Semi-Principled Models}
Following the customization of the Autonomie models, empirical constants and maps were extracted through virtual chassis dynamometer simulations. This process mirrors the concept of testing an actual vehicle on a physical chassis dynamometer. The Autonomie model is subjected to standard EPA drive cycles during simulation. The earlier discussed customizations outlined in \ref{Autonomie Modeling} enable the recording of measurements from specific components of the vehicle, contributing to the extraction of maps and constants. In addition to these empirical values, principled constants and maps were also derived from the base results before any modifications were made to the template Simulink models.

The Autonomie model provides several principled parameters, including $m_{\text{vehicle}}$ (vehicle mass in kg), $m^k_{\text{general}}$ (generalized vehicle mass per gear, accounting for drive-line inertia), $r_{\text{tire}}$ (tire radius in meters), and road load parameters ($R_{\text{a}}$, $R_{\text{r}}$, and $R_{\text{g}}$ representing air resistance, rolling resistance, and frictional load, respectively). Additionally, the model supplies $d_{\text{r}}$ (final drive ratio), $g_{\text{r}}$ (gear ratios), $N_{\text{max}}$ (maximum engine speed in rad/s), and $N_{\text{min}}$ (engine idle speed or minimum engine speed in rad/s). Furthermore, principled maps are directly extracted from Autonomie results data file and stored as arrays for use in the semi-principled model. These maps include $K_{\text{upshift}}(\alpha,v)$ (automatic gear upshift map), $V_{\text{upshift}}(\alpha,k)$ and $V_{\text{downshift}}(\alpha,k)$ (manual upshift and downshift maps), $T_{\text{max}}(N)$ (maximum engine torque map), $T_{\text{wmax}}(v)$ (maximum wheel torque map), and $T_{\text{wmax}}(v,k)$ (maximum wheel torque by gear map). In the semi-principled model, $T_{\text{wmax}}(v,k)$ is employed to calculate the maximum wheel torque and force, while $T_{\text{wmax}}(v)$ is used to determine the pedal angle consistently with Autonomie and to estimate Autonomie's gear choice result when using the upshift map $K_{\text{upshift}}(\alpha,v)$.

Automated procedures are established to extract empirical constants from the Autonomie model. Initially, the Autonomie model undergoes simulation across representative drive cycles~\cite{DriveScheduleEPA} for the specific vehicle under consideration. Key parameters such as the \emph{minimum engine torque} $T_{\text{min}}$ ($\text{Nm}$) and \emph{idling fuel rate} $f_{\text{idle}}$ ($\text{g}/\text{s}$) are determined based on specific conditions, including maintaining a vehicle speed below $0.1\text{m}/\text{s}$ and torque change rate below $0.01\text{Nm}/\text{s}$ within the interval of $[-1\text{s},+1\text{s}]$. The instantiation of fuel-cut involves precise criteria related to speed and wheel force thresholds, independently determined as the $1^\text{st}$ percentile for speed ($v_\text{c}$) and $95^\text{th}$ percentile for wheel force ($F_{wc}$) under specific fuel rate and speed conditions. Downshifting vehicle speeds for the \emph{downshifting map} $K_{\text{downshift}}(v)$ are extracted as median speeds during downshifting events, providing cutoff points for a piece-wise constant mapping. Additionally, the \emph{open torque converter correction for first gear} $T_{\text{correction}}(a)$ ($\text{Nm}$) is determined by averaging acceleration and torque underestimation errors over distinct regions of acceleration. The semi-principled model utilizes fitting routines for Autonomie's VCD data into maps, employing least square fittings of polynomial functions for parameters such as engine speed and engine torque to fuel rate, transmission output speed and wheel force to engine speed for each gear, and transmission output speed and wheel force to engine torque for each gear. These fitting routines are chosen to be of low degree to avoid unnecessary complexity and over-fitting, ensuring consistency across various vehicle classes.

\subsection{Simplified Models}
After constructing the semi-principled models, we proceed with a further fitting step to develop a simplified fuel consumption model for each vehicle class. In this approach, we prioritize model simplicity and interpretability, restricting the simplified model inputs to vehicle speed, acceleration, and road grade. The objective is to create a physics-like and polynomial-spirited model, aiming to avoid optimization problems exploiting overfitting artifacts. The choice of a polynomial function is inspired by physical laws governing the power demand needed to overcome friction forces \cite{Galvin2017, Gong2018}. Consequently, we fit the semi-principled model of each vehicle class to a function $f_{\text{s}}(v,a,\theta)$ of the form 
\begin{equation*}
    f_{\text{s}}(v,a,\theta) = \max\left\{\ell(v,a,\theta), \, f_\text{p}(v,a,\theta)\right\}\;,
\end{equation*}
where
\begin{align*}
\ell(v,a,\theta) &=
  \begin{cases}
        \beta & \text{if $v\leq v_\text{c}$}\;, \\
        0 & \text{if $v> v_\text{c}$ and $a<a_c(v,\theta)$}\;,
  \end{cases}
\quad\text{and} \\
f_\text{p}(v,a,\theta) &=  C(v) + P(v)a + Q(v) (a_+)^2 + Z(v)\theta\;.
\end{align*}

The function $f_\text{p}(v,a,\theta)$ represents fuel rate, and is a polynomial in speed $v$, acceleration $a$, and road grade $\theta$, while $\ell(v,a,\theta)$ is effectively a lower bound on the amount of fuel consumed by the vehicle.
The speed at which a fuel-cut happens is described as the $v_\text{c}$, and the corresponding acceleration values at which fuel-cut is activated is determined by another polynomial function $a_c(v,\theta)$. $C(v)$, $P(v)$ and $Q(v)$ are all polynomial functions of speed which aims to ensure that $f_\text{p}(v,a,0)>0$ is increasing in $a$ by constraining the minimum $f_\text{p}(v,a_{\text{min}}(v),0)$ to be above zero.

For simplicity, we assume that for a fixed speed, the rate at which fuel consumption increases with road grade is independent of vehicle acceleration. The fitting is conducted in an $L^2$ sense, and the implementation involves approximations via quadrature on regular grids. This fitting procedure encompasses the determination of key parameters such as fuel-cut speed ($v_\text{c}$), lower bound parameter ($\beta$), and coefficients for the polynomial terms in the fuel consumption function.
The units associated with the fitted parameters in this convex model across a portfolio of six vehicle classes can be found in \cite{khoudari2023reducing}.

\section{Chassis Dynamometer Experiment}
\subsection{Experiment Setup}
\begin{figure}[htbp]
    \centering 
    \includegraphics[width=0.99\linewidth]{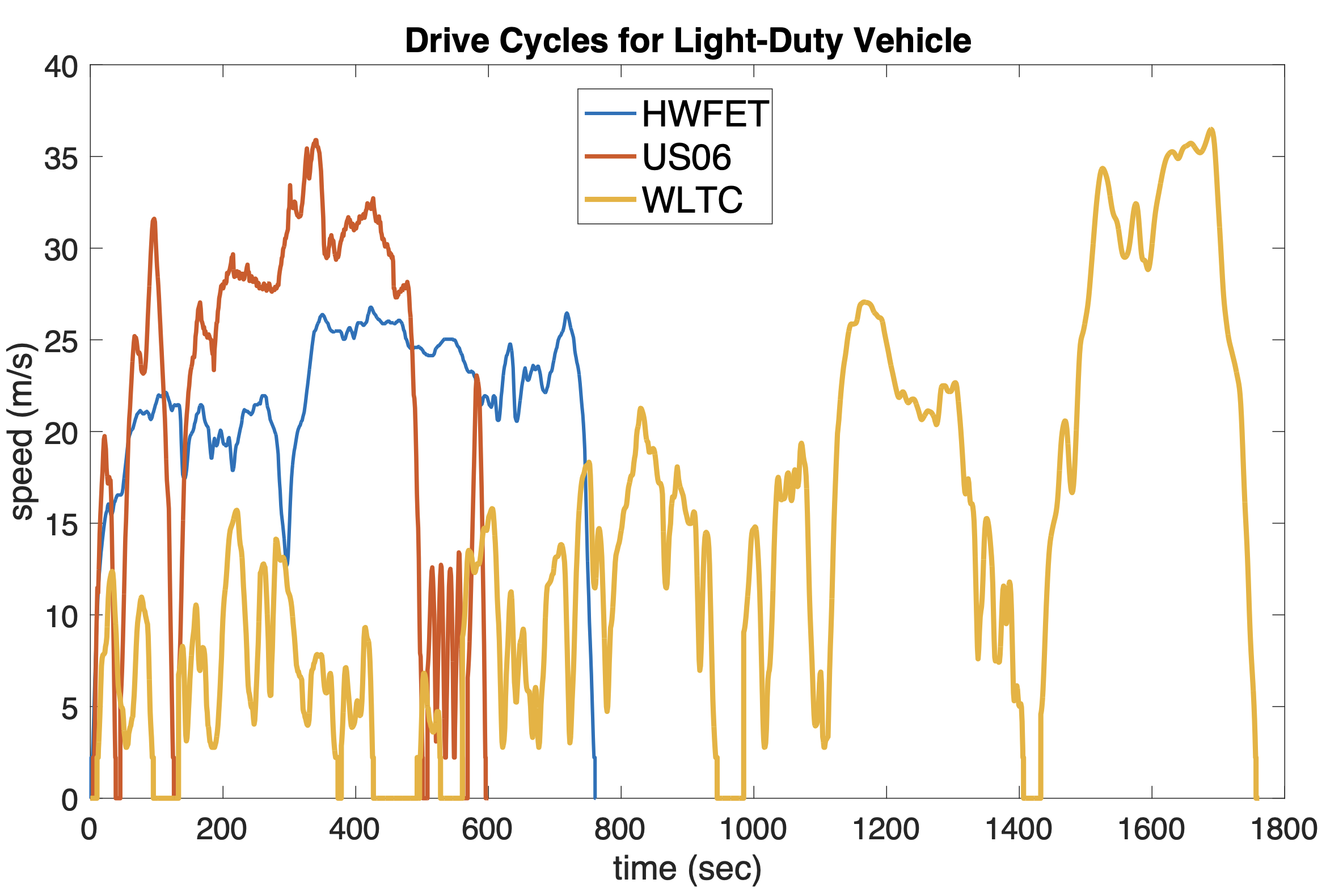}
    \caption{This plot presents the speed profiles of the Standard Drive Cycles used in Light-Duty Vehicle modeling on the Virtual Chassis Dynamometer, which closely mirror those employed in the Real-Vehicle Chassis Dynamometer Experiment.}
    \label{DriceCycles}
\end{figure}

In this experimental phase, our primary objective is to faithfully replicate the driving patterns utilized in the development of the semi-principled model through Autonomie and the virtual chassis dynamometer. The drive cycles employed on the virtual chassis dynamometer, identical to those utilized in the chassis dynamometer experiment, are detailed below and illustrated in Figure \ref{DriceCycles}:
\begin{itemize}
    \item Highway Fuel Economy Test (HWFET) is the highway fuel economy driving schedule for highway driving conditions under 60 mph.
    \item US06 also called the "Supplemental FTP" is a high acceleration aggressive driving schedule.
    \item Worldwide Harmonized Light Vehicles Test Cycle (WLTC) consists of four sequences exploring different levels of speed and acceleration patterns.
\end{itemize}
The listed~\cite{DriveScheduleEPA} driving patterns were deliberately selected to analyze both the transient and cruising behaviors of the vehicle. Each cycle was meticulously executed twice on the chassis dynamometer under hot engine conditions for comprehensive evaluation.

\subsection{Chassis Dynamometer Output Data}
The measurements utilized in our analysis, derived from the Chassis Dynamometer Experiments, include the following:
\begin{itemize}
    \item Speed (kilometer-per-hour)
    \item Engine Speed (revolution-per-minute)
    \item Engine Torque (Newton-meter)
    \item Pedal Angle (\%)
    \item Fuel Rate (grams-per-second)
    \item Engine Water Temperature (degree Celcius)
    \item Gear Selection
    \item Timestamp (second)
\end{itemize}

\begin{figure}[htbp]
    \centering 
    \includegraphics[width=0.99\linewidth]{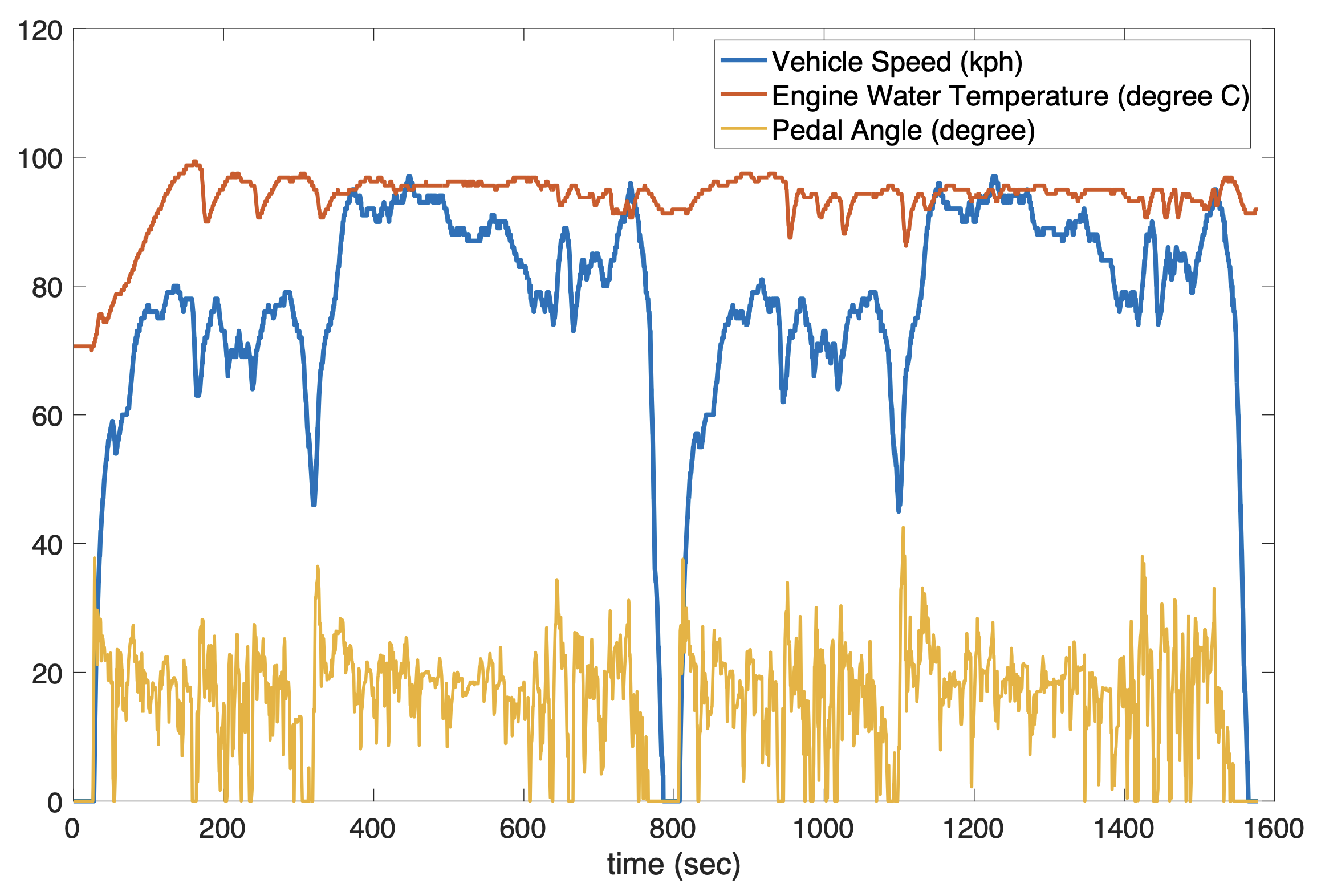}
    \caption{This plot shows the trend of engine water temperature over time as the driver on the chassis dynamometer attempts to execute the target driving pattern. During the first cycle of this example driving pattern, the engine water temperature is still warming up, approaching the "hot mode" condition. }
    \label{EngWaterTemp}
\end{figure}

One initial observation highlights that, despite the diligent efforts of the human operator managing the chassis dynamometer, there are noticeable discrepancies between the two cycles within each set of drive cycles. The primary aim was to conduct a consistency check on the fuel estimates for both cycles, but this proves challenging due to the inherent inconsistencies between them. Consequently, we opted to utilize data from only one of the two cycles. Another compelling reason for this choice is the extended duration during which the engine is not yet in "hot mode" for a significant portion of the first cycle, introducing a potential impact on fuel consumption. This phenomenon is evident in Figure \ref{EngWaterTemp}, illustrating the gradual heating of the engine water temperature within the initial 200 seconds. Such behavior was consistently observed across all other drive cycles during the initiation of the driving pattern. Autonomie assumes the engine is in "hot mode," necessitating the vehicle to also be in such a mode for consistency.

A thorough comparison of the speed profiles used in the chassis dynamometer drive cycles and their counterparts in the virtual chassis dynamometer reveals a notable disparity. In response to this inconsistency, we opted to utilize the recorded speed profile from the chassis dynamometer for evaluating fuel consumption in Autonomie, as well as in the Semi-principled and Simplified models. It is essential to underscore, however, that the speed measurement from the real chassis dynamometer experiment exhibits low resolution, set at 1 kilometer-per-hour (kph), and lacks acceleration data—a crucial parameter for energy models (both the semi-principled and simplified models), given their reliance on acceleration as a key input.

\subsection{Post-processing of Chassis Dynamometer Output Data}
In the post-processing of dynamometer raw data, the primary goal is to produce a speed measurement of elevated resolution and formulate an acceleration profile. These outputs serve as inputs for Autonomie, alongside semi-principled and simplified models. This meticulous approach ensures a uniformity of inputs across all three platforms, fostering consistency and reliability in subsequent analyses.

\begin{figure}[htbp]
    \centering 
    \includegraphics[width=0.99\linewidth]{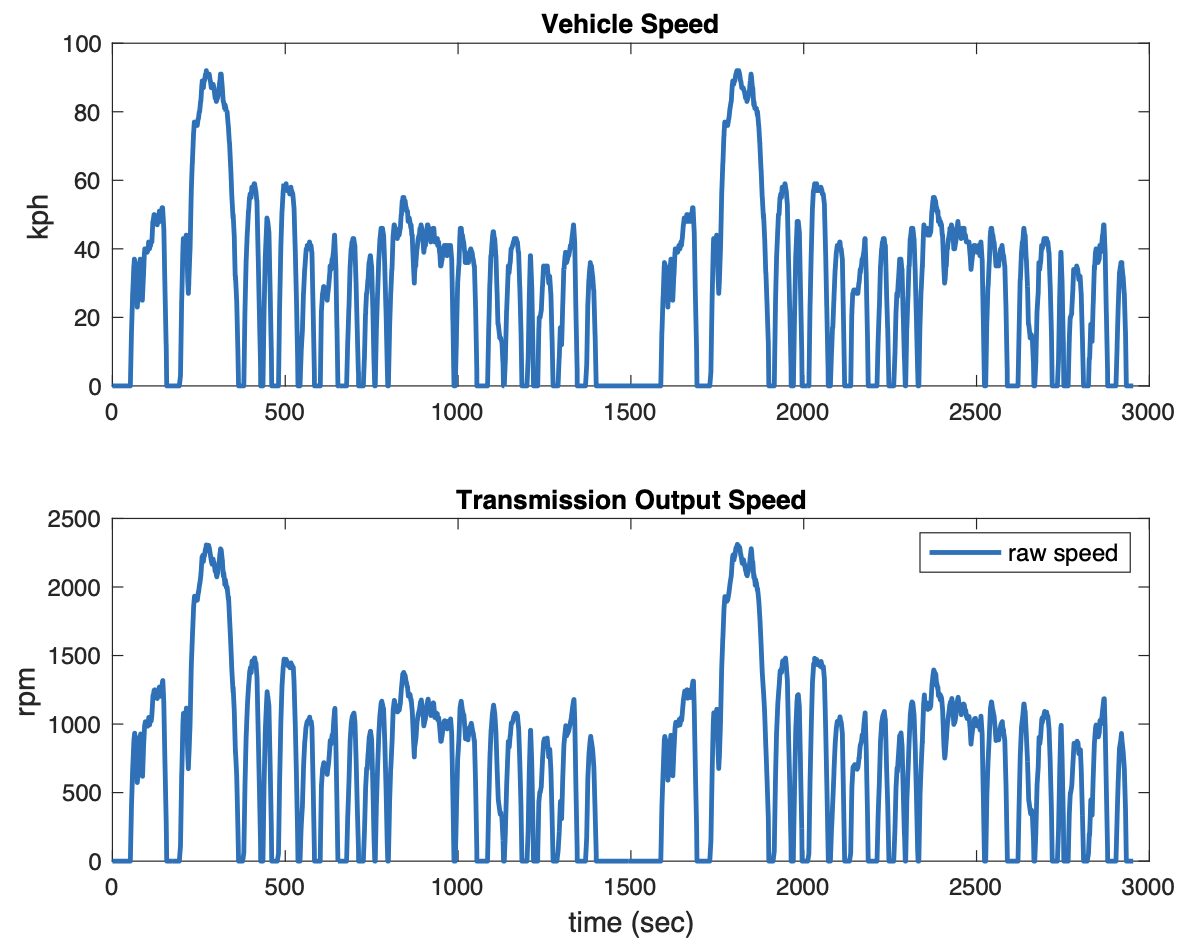}
    \caption{The two plots compare the trends in vehicle speed (in $kph$) and transmission output speed (in $rpm$). This analysis aims to identify potential sources for deriving a higher-resolution speed profile by examining the relationship between these two parameters}
    \label{v_vs_Ntout}
\end{figure}

To obtain a higher resolution of speed profile, we utilized the transmission output speed measurements which show high correlation with the vehicle speed. When plotted with respect to time, the speed (kph) and transmission output speed (rpm) shows identical shapes as shown in Figure \ref{v_vs_Ntout}.

\begin{figure}[htbp]
    \centering 
    \includegraphics[width=0.85\linewidth]{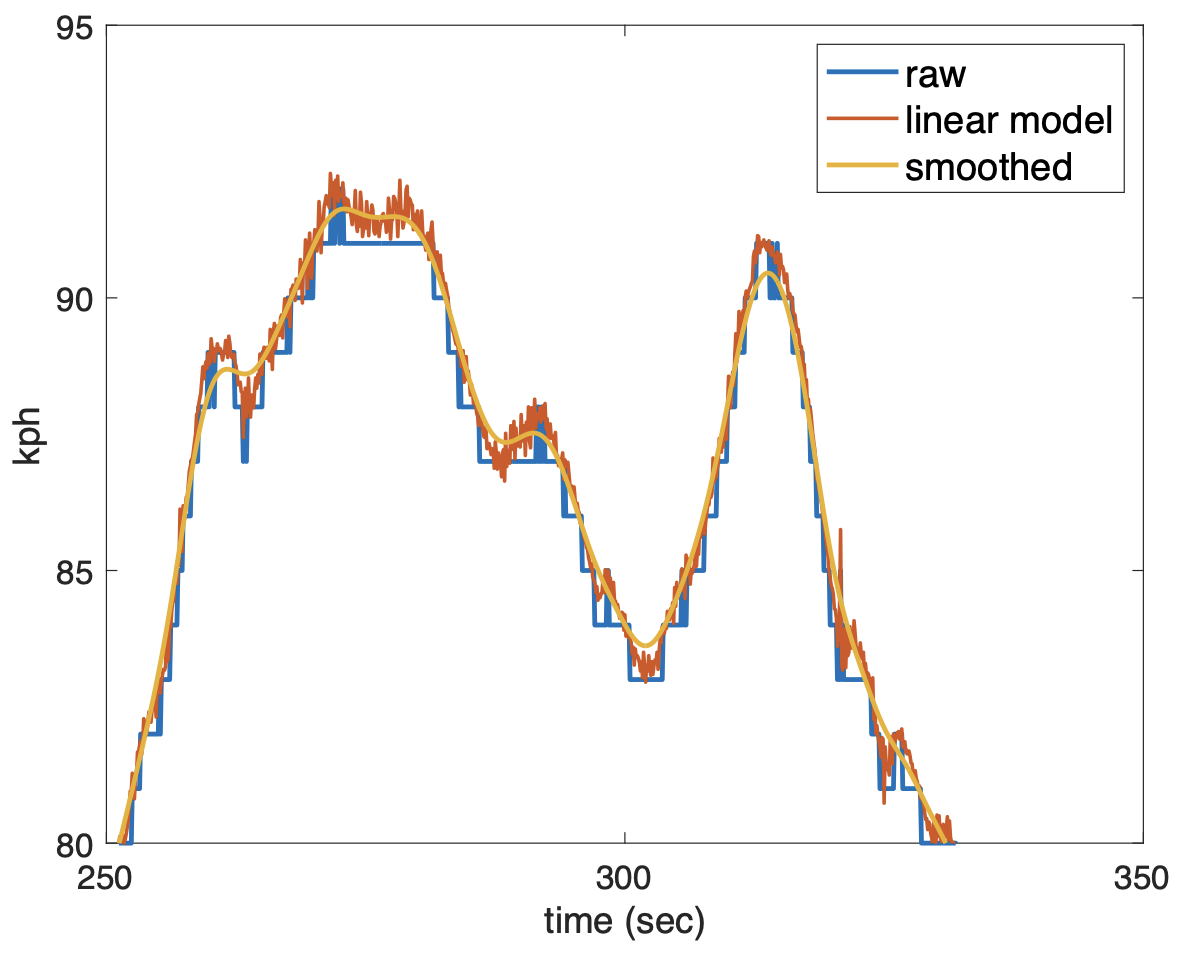}
    \caption{This plot presents a visual examination, comparing the original speed, the speed derived from the transmission output speed using linear regression, and the smoothed version of the derived speed.}
    \label{rescaled_v}
\end{figure}

Using linear regression to derived a model that maps the transmission output speed to the vehicle speed, we obtained the following equation:
\begin{align*}
    v_{\text{derived}} = 0.0398 \times N_{\text{output}}
\end{align*}
The regression model is bounded to a minimum speed of 0 kph. Shown in Figure \ref{rescaled_v} is the comparison between the raw speed data and the linear model zoomed within the 250 sec to 350 sec time window of the WLTC drive cycle.
As illustrated in Figure \ref{rescaled_v}, it is evident that the linear model exhibits superior resolution, albeit with notable noise attributed to the inherent variability in transmission output speed. This noise presents a challenge, particularly since the acceleration profile is directly derived from the speed profile. To address this issue, a mitigation strategy was implemented by applying a smoothing technique to the speed profile. An iterative simple averaging method is used in smoothing the speed profile, and is shown in the equation below.
\begin{align*}
    v_{s} = \frac{\mu}{2}[v_{s}(i-1)] + (1-\mu)[v_{s}(i)]
     + \frac{\mu}{2}[v_{s}(i+1)]
\end{align*}

The index $i$ represents the position of a data point in the sequence of velocity measurements $v_{s}(i)$. For example, $i=1$ corresponds to the first data point, and $i=n$, corresponds to the last data point in the sequence. The smoothing operation is applied to the data points from  $i=2$ to $i=n-1$, ensuring that the smoothed velocity $v_{s}$ at each of these positions is computed based on a weighted average of the current data point and its immediate neighbors. For the boundary data points, the smoothing operation is not applied. Instead, these points remain unchanged: $v_{s}(i) = v_{derived}(i)$ for $i = 1, n$. This equation is initialized with $v_s = v_{derived}$ and $\mu = 0.5$. The $v_s$ is updated $N$ times in iteration using the same equation where $N$ is the number of smoothing steps. From this smoothed speed, the acceleration profile is derived. To filter out outliers in the acceleration measurements, we excluded the upper and lower 5\% of their percentile rank. 

\begin{figure}[htbp]
    \centering 
    \includegraphics[width=0.85\linewidth]{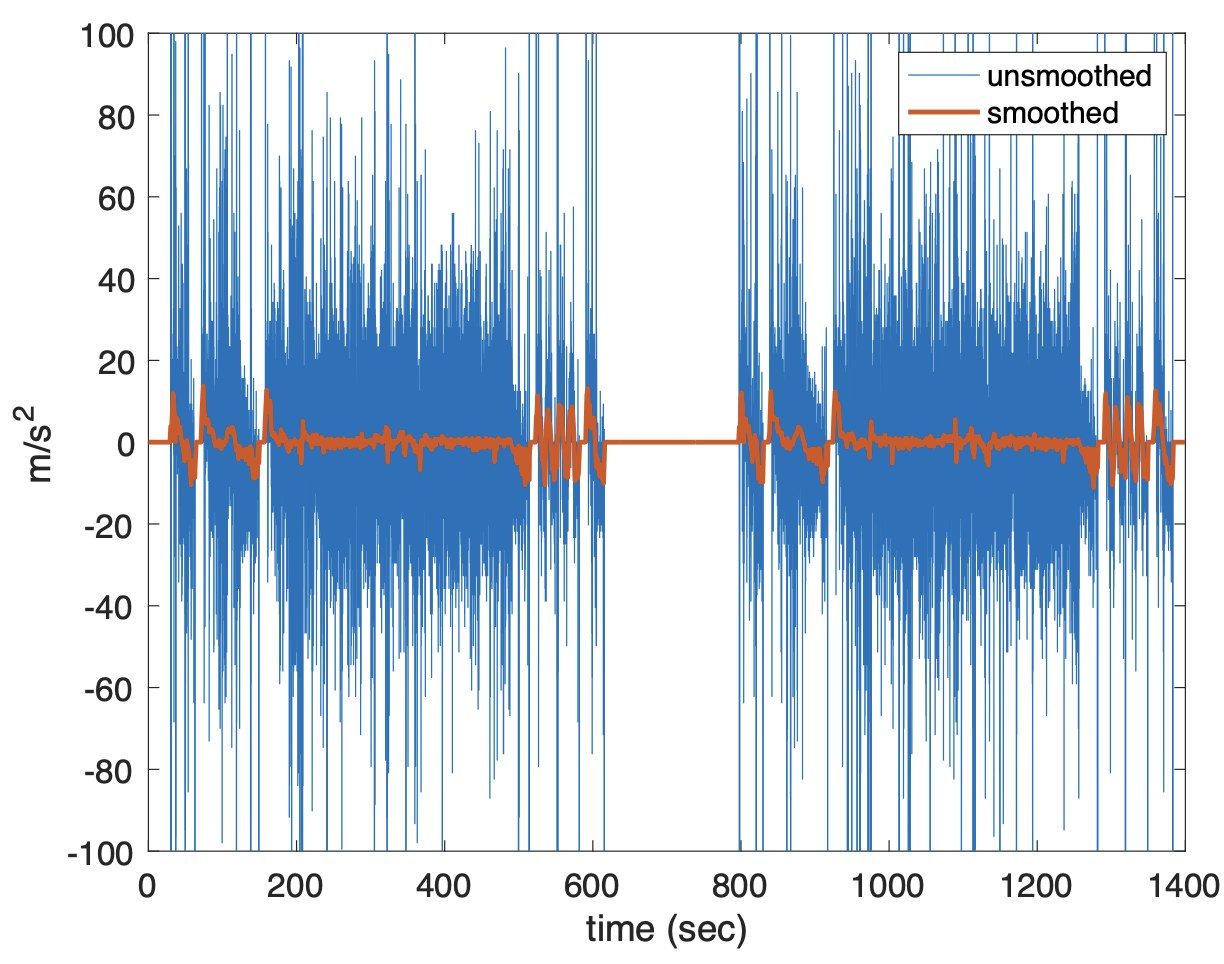}
    \caption{The acceleration value shown are those obtained from the derived speed from the transmission output speed (blue) which is noisy in nature and those derived from the smoothed speed (red). This acceleration profile corresponds to the two cycles of WLTC driving pattern}
    \label{smoothed_accel}
\end{figure}

The determination of the optimal number of smoothing steps is guided by the tracking of improvements in the acceleration bounds. The concluding selection of the number of smoothing steps is made when the enhancements in acceleration converge to a specific value and/or when the acceleration bounds are already within an acceptable range, whereas the acceptable range is set to [$-4 m/s^2$, $4 m/s^2$]. In this context, the chosen number of smoothing steps is not the same for every drive cycles. Some drive cycles are more aggressive than the other making the measurement more prone to noise. The acceleration values are derived as the temporal derivative of the speed profile, expressed as $a = dv/dt$. Figure \ref{smoothed_accel} visually portrays the pronounced enhancement in acceleration following the application of smoothing to the derived speed obtained from the transmission output speed. Notably, prior to the smoothing process, the acceleration values exhibited unrealistic magnitudes, surpassing an absolute value of $100$ $m/s^2$.





\section{Toyota RAV4 Autonomie Model Validation and Calibration}
\label{auto_vs_dyno}

The energy models presented in the energy modeling pipeline in~\cite{khoudari2023reducing}, were designed to align with representative vehicles. These representative vehicles were meticulously selected through a clustering process, ensuring that each representative accurately encapsulates a distinct group of vehicles. This deliberate clustering approach serves to streamline the modeling process by narrowing down the number of vehicles to be modeled.

All vehicles included in the clustering process were the averaged versions of their respective categories, as modeled in Autonomie. One of the cluster representative is the Midsize SUV. It is worth noting that, despite the categorization of the Toyota RAV4 as a Midsize SUV, certain parameters may not precisely match, including but not limited to tire radius, gear-to-axle ratio, road load coefficients, among others. To facilitate a more accurate comparison between the performance of the energy models and the chassis dynamometer output data, specific parameters in the Autonomie model for the Midsize SUV were adjusted to align with the actual characteristics of the Toyota RAV4. Beyond these evident physical characteristics, the execution of the energy modeling adhered to the principles outlined in \cite{khoudari2023reducing}.

In this section, we seek to show the key parameters in the Autonomie model for Toyota RAV4 that were calibrated, starting from the MidSUV template model. Then, we validated the resulting Autonomie model with the actual dynamometer data.


\subsection{Validation Metrics}
For the purpose of performance comparison, we focused on the following parameters:

\noindent\textbf{Fuel Rate}: The fuel rates were obtained by running the Autonomie (as well as the semi-principled, and simplified models) using the smoothed speed data from the dynamometer output and the corresponding derived acceleration.

\noindent\textbf{Cumulative Fuel}: Cumulative fuel represents the total fuel consumption at any given point in time, with the final value reflecting the total fuel consumed over the entire drive cyle.

\noindent\textbf{Internal Vehicle Dynamics}: This category includes parameters such as Engine Speed (revolutions per minute), Engine Torque (Newton-meters), Pedal Angle (percentage), and Gear Schedule.

\subsection{Autonomie Model Calibration using Dynamometer Data}
\begin{figure*}[htbp]
    \centering 
    \includegraphics[width=0.99\linewidth]{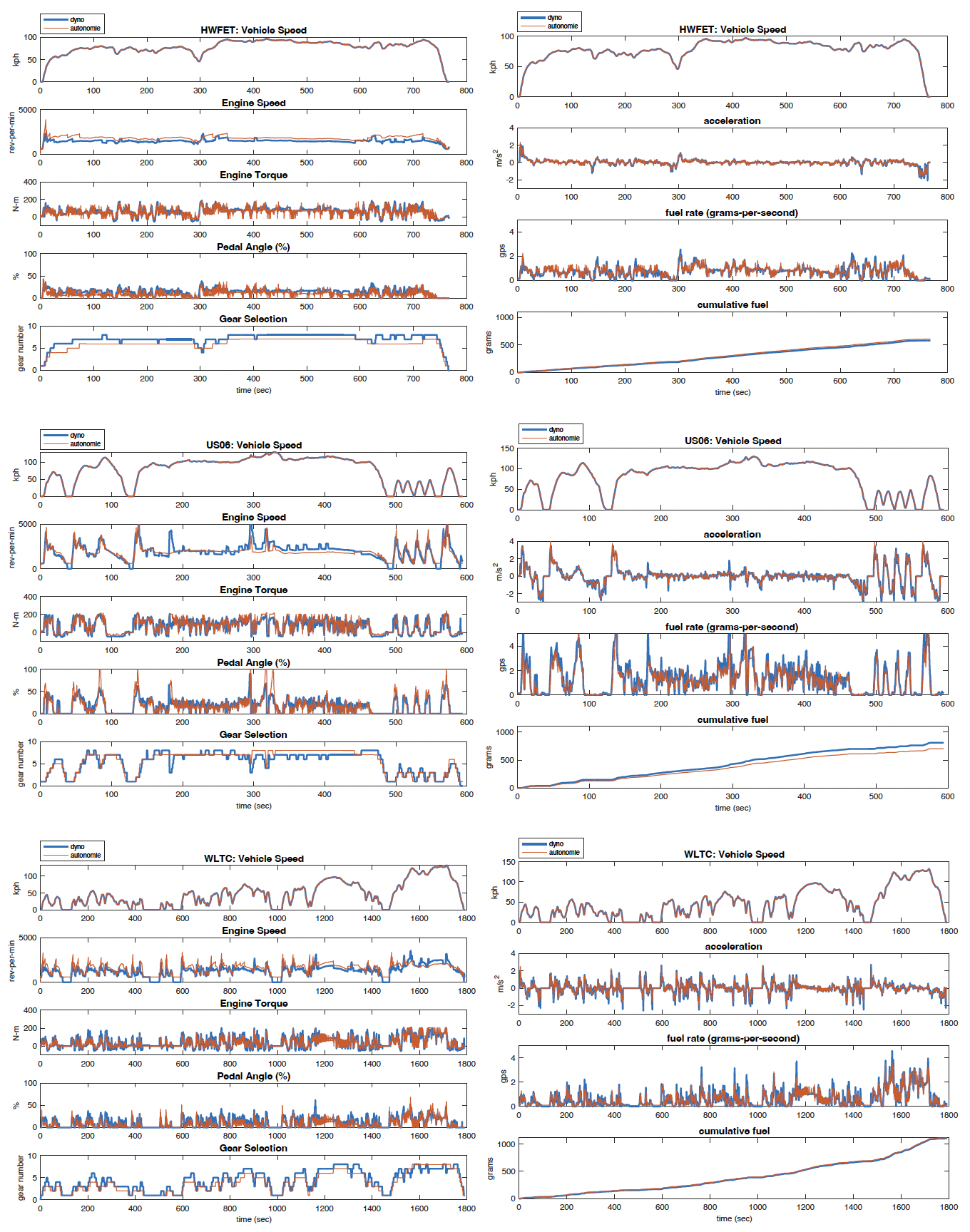}
    \caption{This comparison illustrates the internal dynamics (left column) and fuel estimation (right column) performance of the baseline Autonomie model for the Toyota RAV4, which was derived from the MidSUV template model in Autonomie, against the measurements from an actual Toyota RAV4 on a chassis dynamometer. The plots presented cover both cruising (HWFET) and transient (WLTC and US06) driving patterns.}
    \label{dynamics comparison before}
\end{figure*}

As a starting point, we selected a template Autonomie model that closely aligns with the characteristics of a Toyota RAV4, specifically in the Midsize SUV category. Subsequently, we fine-tuned the template model to mirror the specifications of the Toyota RAV4, referencing publicly available records. Fortunately, Toyota RAV4 is one the vehicles that was benchmarked by Environmental Protection Agency (EPA) in assessing the effectiveness of advanced low emission and low fuel consumption technologies for a broad range of key light-duty vehicles~\cite{ToyotaEngineBenchmark}, making its engine performance publicly available. Despite limited resources detailing the specific parameters requiring adjustment, we made initial modifications based on the information at hand. The complete procedure on how to modify the Autonomie template model is highlighted in \cite{khoudari2023reducing}. The comparison between the the dynamometer data and the Autonomie model output for the same driving patterns after these initial adjustments is illustrated in Figure \ref{dynamics comparison before}.

\begin{figure*}[htbp]
    \centering 
    \includegraphics[width=0.99\linewidth]{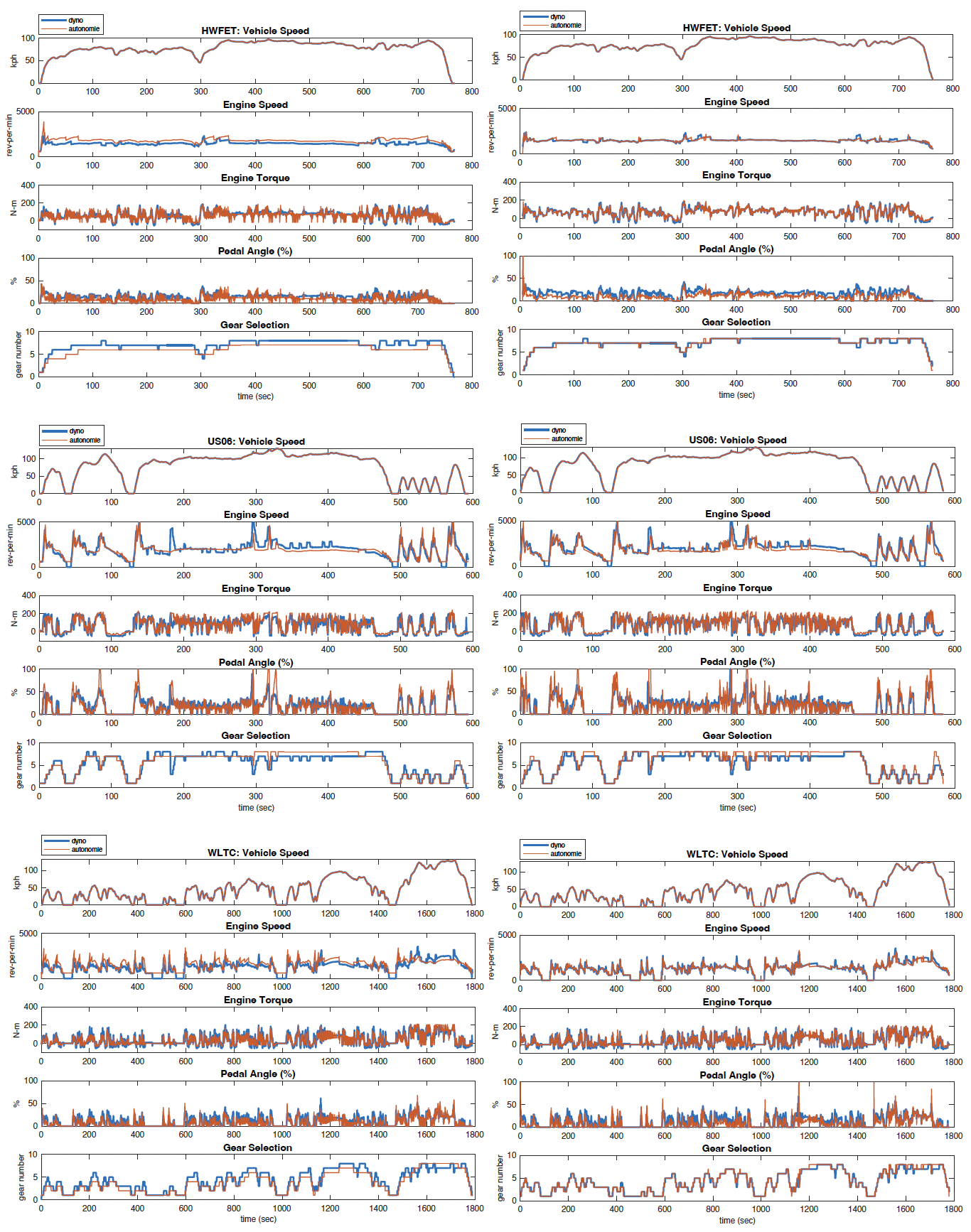}
    \caption{These plots compare the internal dynamics performance between the \emph{baseline} (left column) and \emph{improved} (right column) Autonomie models for the Toyota RAV4 against measurements obtained from an actual Toyota RAV4 on a chassis dynamometer during four drive cycles. Notably, gear scheduling and engine speed have shown significant improvements across all three drive cycles. The engine torque in the improved model is substantially better for the HWFET cycle, although it does not perform as well in the other two cycles. After calibration, the pedal angle is reduced across all drive cycles, except in some sections in the WLTC cycle, where significant spikes are observed.}
    \label{dynamics comparison after}
\end{figure*}

In this comparison, a noticeable observation is that Autonomie consistently upshifts earlier than the dynamometer, as shown in the gear schedule plots for all drive cycles in Figure \ref{dynamics comparison before}. Another key observation is that the baseline Autonomie model overestimates engine speed during cruising, although it performs comparatively better for transient driving patterns, with the caveat that it fails to reach zero engine speed. The engine torque values for both cruising and transient driving patterns closely align with the dynamometer data, except that the model falls short of reaching peak torque levels. Notably, the Autonomie model consistently registers a lower pedal angle compared to the actions executed by the dynamometer throughout most of the time. This could introduce a minor discrepancy in fuel rate estimation. While seemingly insignificant in individual instances, the cumulative effect could be substantial. To address this issue, we made further adjustments to the gear scheduling method of the Autonomie model, resulting in significant improvements in gear selection and engine torque for both transient and cruising driving patterns, as shown in Figure \ref{dynamics comparison after}. Although the engine speed has not changed significantly numerically, the plots clearly demonstrate a higher precision in the majority of data points. It is also important to emphasize that modifications made in the baseline Autonomie models did not make that much impact on the US06 cycle, as the numerical result suggest no significant difference for the baseline and the improved Autonomie models.

\begin{figure*}[htbp]
    \centering 
    \includegraphics[width=0.99\linewidth]{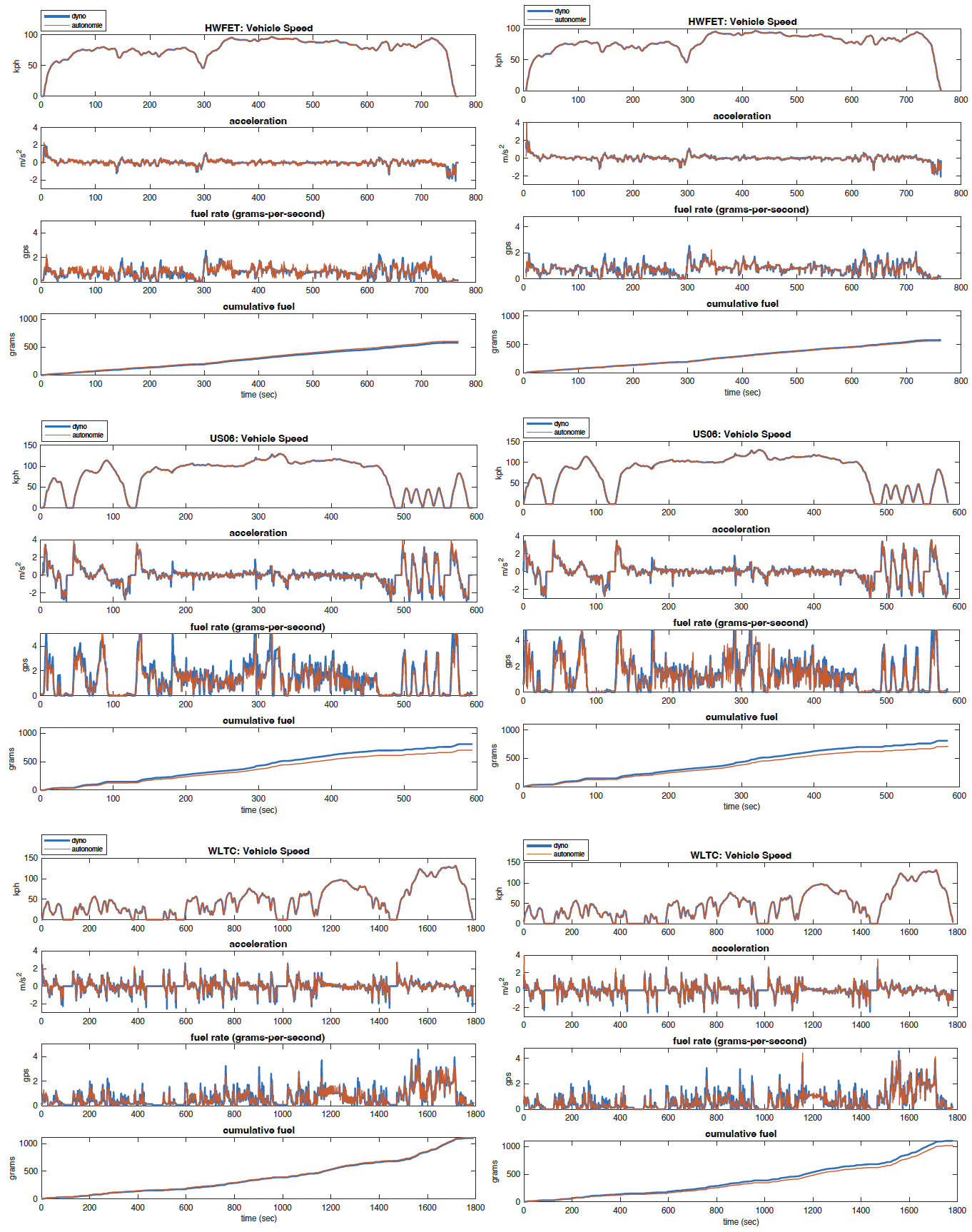}
    \caption{These plots compare the fuel rate and cumulative fuel consumption between the baseline (left column) and improved (right column) versions of the Autonomie model against dynamometer data for both cruising (HWFET) and transient (US06 and WLTC) driving patterns. While the fuel estimates, both cumulative and rate, show significant improvement in the HWFET cycle, they have unexpectedly worsened in both transient driving patterns.}
    \label{dyno_vs_autonomie}
\end{figure*}

Figure \ref{dyno_vs_autonomie} present comparisons of fuel rate and cumulative fuel for the dynamometer data versus the two Autonomie models for the Toyota RAV4—both the initial model and the enhanced version.
From the presented figures, it is evident that both versions of the autonomie model faithfully execute the speed derived from the dynamometer data, resulting in comparable fuel consumptions for identical input driving patterns. Notably, the Autonomie model tends to overestimate the fuel rate for highway driving patterns (HWFET) and underestimate it for more aggressive driving patterns, such as the US06 and WLTC. The improved Autonomie model demonstrates enhanced performance for HWFET but exhibits a worse underestimation for aggressive driving patterns than the baseline Autonomie model.

To thoroughly evaluate the estimation accuracy of the Autonomie model, we calculated the mean absolute error (MAE) across various aspects, including internal dynamics, moment-by-moment fuel rate, and cumulative fuel consumption. This metric offers valuable insights into the quality of the model's estimates, allowing for a nuanced analysis of its performance across different speed profiles. By utilizing the MAE, we can identify areas where the model excels and where it falls short, providing a detailed understanding of its strengths and weaknesses. Additionally, we examined instances where there was a mismatch in gear scheduling between the dynamometer and the Autonomie model, calculating the percentage of occurrences for these discrepancies.

As shown in Tables \ref{table:auto1_vs_dyno_mae} and \ref{table:auto2_vs_dyno_mae}, the baseline model exhibits slightly better accuracy in engine speed and pedal angle, while the updated model demonstrates significant improvements in engine torque across all three drive cycles. However, gear scheduling has markedly improved in the updated model for the HWFET and WLTC cycles, with almost no change observed for the US06 cycle. Although the goal of enhancing gear scheduling was achieved, it did not automatically translate into better pedal angle performance, which shows only minimal changes across all three drive cycles.

Table \ref{table:auto_vs_dyno_mae} summarizes the differences in cumulative fuel consumption and the MAE for moment-by-moment fuel rate measurements between both Autonomie models and the reference dynamometer data. Notably, the improved model exhibits a smaller MAE in moment-by-moment fuel rate estimation for transient driving patterns compared to the baseline Autonomie model, although it has a larger cumulative fuel error, particularly for the WLTC cycle. In contrast, the cruising driving pattern shows significant improvements in both cumulative and moment-by-moment fuel estimates for the improved Autonomie model.

\begin{table}[htbp]
\caption{Error in Internal Dynamics observed during both cruising and transient driving patterns of the \emph{Baseline} Autonomie Model versus the Dynamometer Data.}
\label{table:auto1_vs_dyno_mae}
\begin{center}
\begin{tabular}{|l|l|l|l|l|l|}
\hline
& HWFET & WLTC  & US06 \\
\hline
\hline
MAE in Engine Speed (rpm) & 1320 & 1177 & 1802\\
\hline
MAE in Engine Torque (Nm) & 26.0554 & 19.9363 & 26.9569\\
\hline
MAE in pedal angle (\%) & 6.2583 & 4.2941 & 6.1735\\
\hline
MAE in gear selection & 1.0534 & 0.7409 & 0.7103\\
\hline
mismatch in gear selection (\%) & 15.30 & 9.48& 10.60\\
\hline
\end{tabular}
\label{table:HWFET_fr_fc_comparison}
\end{center}
\end{table}

\begin{table}[htbp]
\caption{Error in Internal Dynamics observed during both cruising and transient driving patterns of the \emph{Improved} Autonomie Model versus the Dynamometer Data.}
\label{table:auto2_vs_dyno_mae}
\begin{center}
\begin{tabular}{|l|l|l|l|l|l|}
\hline
& HWFET & WLTC  & US06 \\
\hline
\hline
MAE in Engine Speed (rpm) & 1274 & 1154 & 1772\\
\hline
MAE in Engine Torque (Nm) & 11.5112 & 13.8256 & 24.8892\\
\hline
MAE in pedal angle (\%) & 6.9673 & 4.7344 & 6.9621\\
\hline
MAE in gear selection & 0.1345 & 0.2582 & 0.7378\\
\hline
mismatch in gear selection (\%) & 0.39 & 1.63& 10.82\\
\hline
\end{tabular}
\label{table:HWFET_fr_fc_comparison}
\end{center}
\end{table}

\begin{table}[htbp]
\caption{Error in fuel rate and total fuel consumption observed during both cruising and transient driving patterns of Baseline and Improved Autonomie Model versus Dynammeter Data.}
\label{table:auto_vs_dyno_mae}
\begin{center}
\begin{tabular}{|l|l|l|l|l|l|}
\hline
& HWFET & WLTC  & US06 \\
\hline
\hline
Baseline Autonomie & 0.1482 & 0.1339 & 0.3329\\
MAE in fuel rate (gps) &  & &  \\
\hline
Baseline Autonomie & 4.64 & 0.94 & 13.15\\
cumulative fuel error (\%) & & & \\
\hline
\hline
Improved Autonomie & 0.0773 & 0.1024 & 0.2999\\
MAE in fuel rate (gps) & & &\\
\hline
Improved Autonomie & 2.90 & 7.63 & 12.50\\
cumulative fuel error (\%) & & & \\
\hline
\end{tabular}
\label{table:HWFET_fr_fc_comparison}
\end{center}
\end{table}


While substantial efforts have been dedicated to enhancing the internal dynamics performance of the improved Autonomie model, it is crucial to underscore that the fuel rate estimation does not exhibit a notable improvement. The pursuit of further refinements remains viable, acknowledging that certain adjustments may not be immediately evident. It is imperative to emphasize that, at this juncture, the improved model is considered the final autonomie model for the Toyota RAV4. From this finalized model, we derive semi-principled and simplified energy models.

\section{Semi-principled and Simplified Models Validation with Dynamometer Data}
After finalizing the Autonomie model for the Toyota RAV4, we proceeded to derive the semi-principled and simplified models utilizing the methodology outlined in~\cite{khoudari2023reducing} and briefly discussed in Section \ref{section:energy models} of this paper. Employing a consistent pipeline, akin to the Autonomie and dynamometer comparison illustrated in Section \ref{auto_vs_dyno}, we evaluated the fuel rate estimates of both the semi-principled and simplified models. This evaluation utilized the same speed and acceleration profile extracted from the dynamometer data, which was also used for evaluating fuel rate in the Autonomie model.

\begin{figure*}[htbp]
    \centering 
    \includegraphics[width=0.99\linewidth]{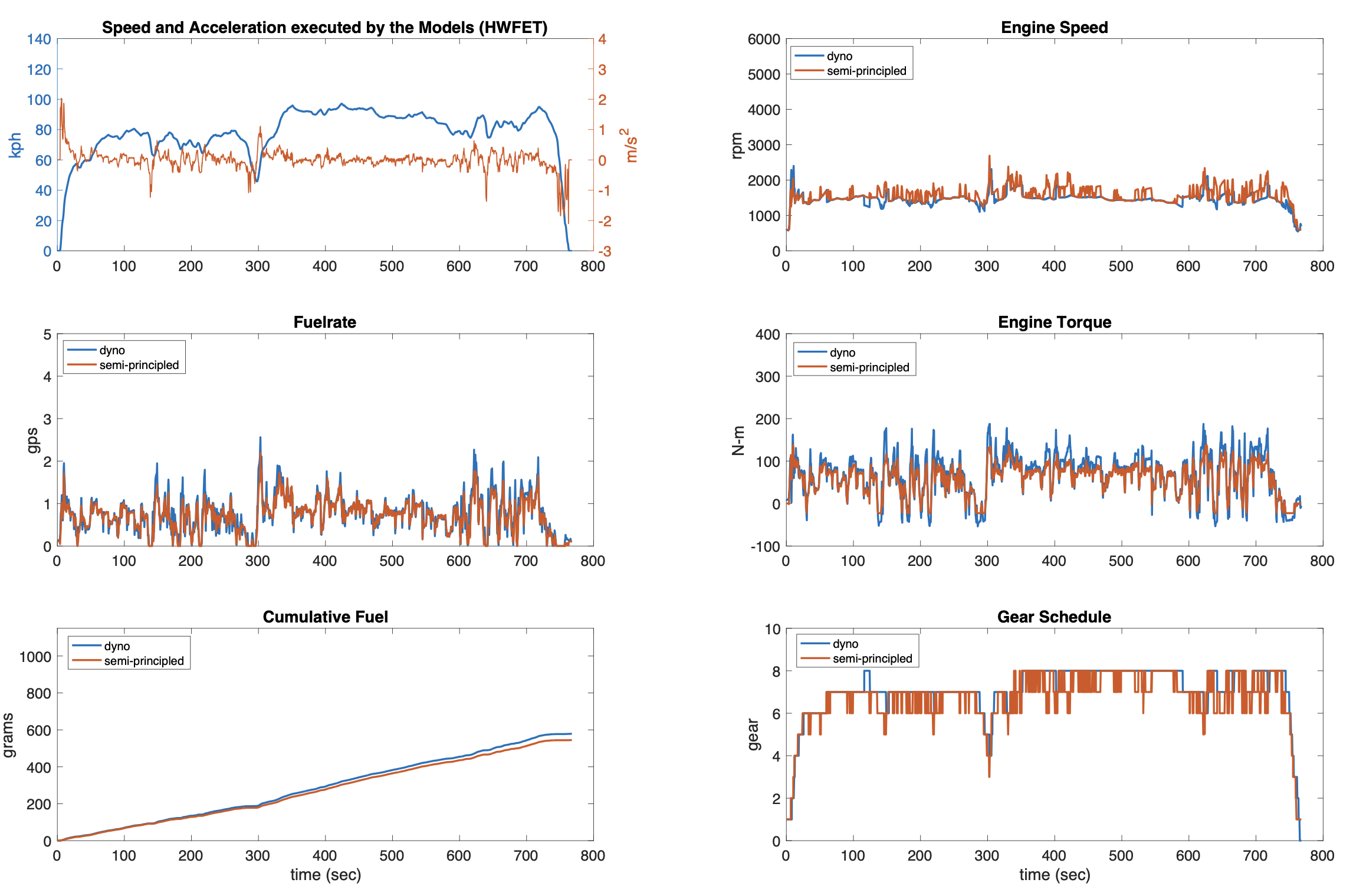}
    \caption{These plots compare the semi-principled model with dynamometer data for the cruising driving pattern (HWFET). While the engine speed in the semi-principled model generally follows the same trend as the dynamometer data, there are random spikes during sections where the vehicle is decelerating. Additionally, the gear scheduling in the semi-principled model shows quick downshifts at moments when the car is decelerating.}
    \label{dyno_vs_semi_hwfet}
\end{figure*}

\begin{figure*}[htbp]
    \centering 
    \includegraphics[width=0.99\linewidth]{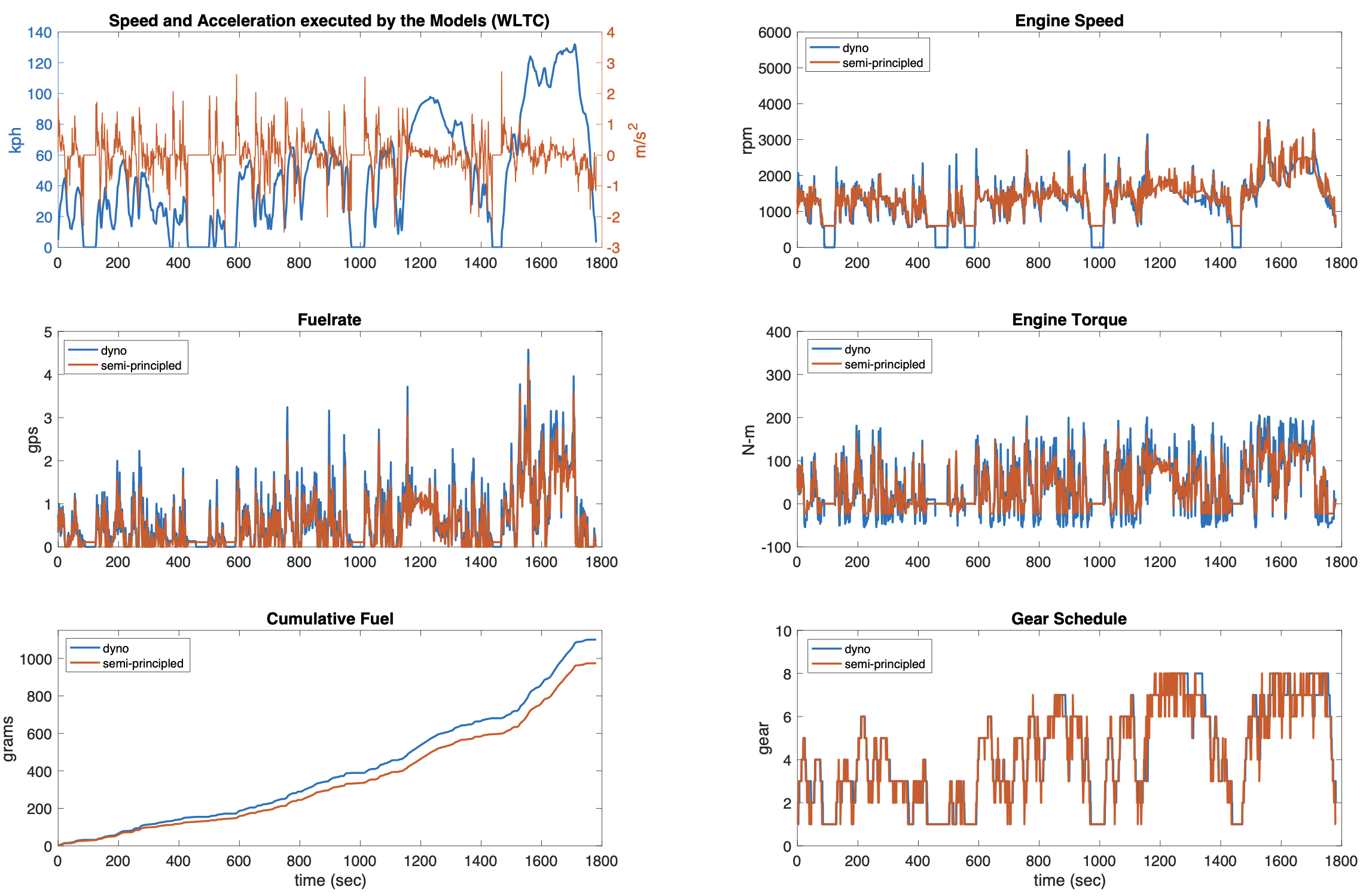}
    \caption{These plots compare the semi-principled model with dynamometer data for the WLTC, a transient driving pattern. While the gear scheduling aligns well with the dynamometer data for the majority of data points, the engine speed and engine torque fail to reach their peak values, leading to underestimation for most of the time. Additionally, the engine speed does not drop to zero when the car is idle, which results in a non-zero fuel rate during these idle moments. This discrepancy has significantly impacted the total fuel consumption.}
    \label{dyno_vs_semi_wltc}
\end{figure*}

Figures \ref{dyno_vs_semi_hwfet} and \ref{dyno_vs_semi_wltc} present comparisons of fuel rate and cumulative fuel consumption between the semi-principled model and the dynamometer data for both cruising and transient driving patterns, respectively. The performance of the derived Autonomie model, when benchmarked against the dynamometer data, shows that the HWFET cycle exhibits a lower error compared to the WLTC and US06 cycles. Table \ref{table:dyno_vs_semi} provides a numerical comparison between the internal dynamics and fuel consumption predicted by the semi-principled model and the actual measurements from the dynamometer. While the semi-principled model is slightly less accurate than the root Autonomie model, the difference is minimal. Importantly, the Autonomie and semi-principled models exhibit a consistent relationship: the better the Autonomie model performs for a specific driving pattern, the more accurate the corresponding semi-principled model is likely to be for that same driving pattern.

\begin{table}[htbp]
\caption{Performance comparison between Toyota RAV4's Dynamometer measurements and Semi-principled Model}
\label{table:dyno_vs_semi}
\begin{center}
\begin{tabular}{|l|l|l|l|l|l|}
\hline
& HWFET & WLTC  & US06 \\
\hline
\hline
MAE in fuel rate (gps) & 0.0812 & 0.1233 & 0.2932 \\
\hline
Cumulative fuel error (\%)& 5.13 & 12.3 & 18.39 \\
\hline
MAE in Engine Speed (rpm) & 1300 & 1175 & 1179 \\
\hline
MAE in Engine Torque (Nm) & 14.7456 & 13.1256 & 22.4815 \\
\hline
MAE in gear scheduling & 0.4352 & 0.3736 & 0.7105 \\
\hline
Mismatch in Gear Selection (\%) & 3.55 & 2.49 & 3.91 \\
\hline
\end{tabular}
\end{center}
\end{table}

\begin{figure*}[htbp]
    \centering 
    \includegraphics[width=0.99\linewidth]{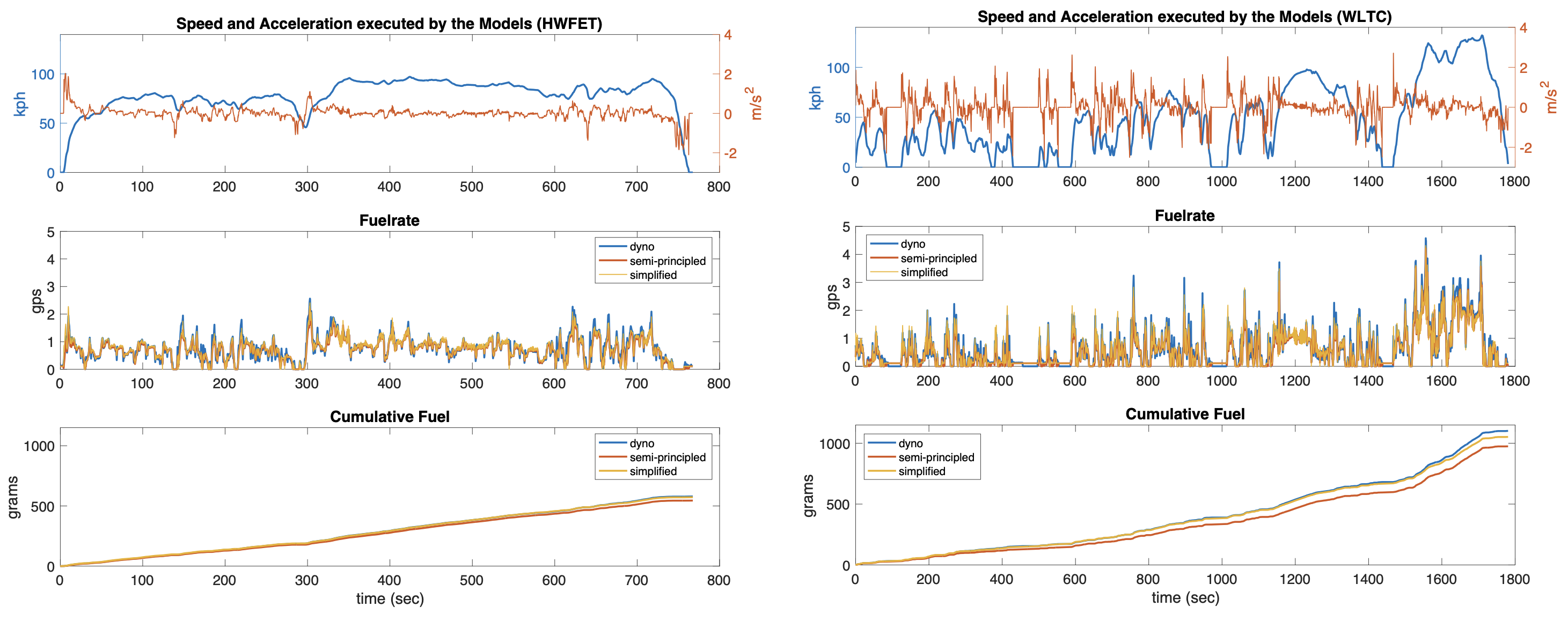}
    \caption{These plots compare the semi-principled model and simplified models with dynamometer data for both cruising (HWFET) and transient (WLTC) driving patterns. The semi-principled model is known for its tendency to underestimate fuel rate consumption. This issue is addressed in the simplified models, as highlighted in \cite{khoudari2023reducing}. The plots demonstrate that the simplified models provide fuel rate estimates that more closely align with the actual vehicle data for both drive cycles.}
    \label{all_models}
\end{figure*}

In both cruising and transient driving patterns, the semi-principled and simplified models consistently underestimate the moment-by-moment fuel rate, resulting in lower cumulative fuel consumption, as shown in Figures \ref{dyno_vs_semi_hwfet} and \ref{dyno_vs_semi_wltc}. A key observation from the semi-principled model's performance is the occurrence of spikes in engine speed during deceleration in the HWFET drive cycle, although the overall trend in engine speed remains relatively similar to the dynamometer data. These spikes were not observed in the WLTC cycle. Additionally, both cruising and transient drives exhibit peculiar artifacts, where the peak and minimum values for engine speed and torque are never fully reached by the semi-principled model. Regarding gear scheduling, there are instances of quick downshifts at certain data points, typically by only one gear. However, overall, the gear schedule of the semi-principled model aligns with the dynamometer data for the majority of the time.

Then, the simplified model was derived from the semi-principled model. To evaluate its fuel rate and cumulative fuel consumption, the same derived speed and acceleration from the dynamometer data were used. Table \ref{table:dyno_vs_sim} summarizes the fuel estimation performance of the simplified models across all three drive cycles. As shown, the simplified model outperforms the semi-principled model, addressing the underestimation issue highlighted in \cite{khoudari2023reducing}. However, it is important to note that despite the overall better performance of the simplified model, its fuel estimate for the US06 cycle does not show significant improvement compared to the improved Autonomie model and the semi-principled model. This indicates that the energy modeling pipeline struggles to adapt effectively to fast-changing acceleration patterns, such as those encountered in the US06 drive cycle. 

\begin{table}[htbp]
\caption{Performance comparison between Toyota RAV4's Dynamometer measurements and Simplified Model}
\label{table:dyno_vs_models_semi}
\begin{center}
\begin{tabular}{|l|l|l|l|l|l|}
\hline
& HWFET & WLTC  & US06 \\
\hline
\hline
MAE in fuel rate (gps) & 0.0904 & 0.1194 & 0.2699 \\
\hline
Cumulative fuel error (\%)& 0.57 & 2.5 & 14.07 \\
\hline
\end{tabular}
\label{table:dyno_vs_sim}
\end{center}
\end{table}

Further optimization of the estimation performance of the Autonomie model for the Toyota RAV4 could be achieved with additional references. However, it's important to note that limited resources are currently constraining this optimization effort.

\section{Conclusion}
In conclusion, our study focused on the Chassis Dynamometer Experiment to replicate driving patterns using Autonomie and a Virtual Chassis Dynamometer. We conducted extensive experiments with varied drive cycles, aiming to analyze both transient and cruising vehicle behaviors. This also required us to do post-processing of dynamometer data which involved improving the speed profile's resolution and formulating an acceleration profile, crucial for energy models, to pave way for uniform inputs for the Autonomie models, the Semi-principled and the Simplified model. Certainly, ensuring alignment between the Chassis Dynamometer trajectory and the inputs for both the Autonomie and derived mathematical models is a commendable achievement and the best feasible solution given the circumstances. The meticulous efforts in post-processing to enhance speed profile resolution contribute to this alignment. However, it is acknowledged that the noisy nature of the derived enhanced-resolution dynamometer speed introduces a level of uncertainty. Any slight discrepancy in speed may propagate into the derived acceleration profile, potentially influencing fuel rate estimations, albeit not necessarily significantly. The study's transparency about this limitation is crucial for a nuanced interpretation of the results, highlighting the importance of minimizing uncertainties in the speed profile for more precise model inputs.

The calibration of the Autonomie model for Toyota RAV4 was a meticulous process. We began with a Midsize SUV template model, fine-tuning it to match RAV4 specifications. The comparison revealed that the initial Autonomie model exhibited some discrepancies, leading to further modifications and an improved model using the dynamometer output measurements. The enhanced Autonomie model demonstrated superior performance over the baseline, particularly in terms of gear scheduling and precision in engine speed when compared to the Dynamometer output measurements. But despite achieving improved internal dynamics performance, the fuel rate estimation did not show significant improvement. In fact, it even shows worse estimate for WLTC and almost no effect for US06.

The improved model, which is considered the final version, was used to derive the semi-principled and simplified models. Comparisons with dynamometer data revealed precise fuel estimations, although there was a slight tendency to overestimate. The moment-by-moment fuel rate difference between the Autonomie model and the Semi-principled and Simplified models remained within 0.07 grams per second, indicating a strong level of consistency. This suggests that the precision of the energy models is closely linked to the accuracy of the root Autonomie model from which they were derived. As long as the initial Autonomie model accurately replicates real vehicle data, the derived energy models can achieve a similar level of precision. 


Further refinements and optimizations may be possible, acknowledging potential adjustments not immediately evident. The study's current status considers the improved Autonomie model final for the Toyota RAV4, from which semi-principled and simplified energy models are derived. Future work may involve additional references and optimization efforts for a better Autonomie model. Indeed, the validation results clearly show that both the semi-principled and simplified models provide realistic estimates of fuel rate consumption on flat roads. Despite some overestimation, the models consistently capture the behavior of the actual vehicle, emphasizing their reliability in predicting fuel consumption patterns.

\section*{Acknowledgment}
The authors would like to thank: (a) the Autonomie team (Argonne National Laboratory), particularly Sylvain Pagerit, Aymeric Rousseau, and Ram Vijayagopal for suggestions and guidance in the usage of the Autonomie software; (b) Sean Murphy (Toyota) for helpful suggestions on the model fitting; (c) Mike Huang (formerly of Toyota) for developing and contributing the initial structure of the semi-principled models used herein and (d) all members of the CIRCLES team who provided helpful feedback on the created models. This material is based upon work supported by the U.S.\ Department of Energy’s Office of Energy Efficiency and Renewable Energy (EERE) under the Vehicle Technologies Office award number CID DE--EE0008872. The views expressed herein do not necessarily represent the views of the U.S.\ Department of Energy or the United States Government. Research was supported by National University - Manila, Philippines (Joy Carpio) and King Abdulaziz City for Science and Technology (S.~Almatrudi). The views and conclusions contained in this document are those of the authors and should not be interpreted as representing the official policies, either expressed or implied, of the Army Research Office or the U.S. Government. The U.S. Government is authorized to reproduce and distribute reprints for Government purposes notwithstanding any copyright notation herein.


\begin{thebibliography}{99}

\bibitem{CIRCLES_AMR2020}
A.~M.~Bayen, J.~W.~Lee, B.~Piccoli, B.~Seibold, J.~M.~Sprinkle, and D.~B.~Work,
\newblock "CIRCLES: Congestion Impacts Reduction via CAV-in-the-loop Lagrangian Energy Smoothing,"
\newblock Presented at the 2020 Vehicle Technologies Office Annual Merit Review, Washington, DC (virtual), 2020.

\bibitem{bhadani2018dissipation}
Rahul~Kumar~Bhadani, Benedetto~Piccoli, Benjamin~Seibold, Jonathan~Sprinkle, and Daniel~Work,
\newblock "Dissipation of emergent traffic waves in stop-and-go traffic using a supervisory controller,"
\newblock In \emph{2018 IEEE Conference on Decision and Control (CDC)}, pp. 3628--3633. IEEE, 2018.

\bibitem{carpio2018traffic}
J.~N.~Carpio, J.~J.~Guadalupe, K.~M.~Ilano, J.~Ranada, C.~L.~Landrito, J.~D.~Farala, M.~Bagara, A.~M.~Hilado, and M.~R.~Doringo,
\newblock "Traffic congestion and speed assessment using image processing technology accessible via internet through smart devices,"
\newblock \emph{International Journal of Simulation: Systems, Science \& Technology}, vol. 19, no. 3, pp. 9.1--9.6, June, 2018.

\bibitem{chou2022lord}
Fang-Chieh~Chou, Alben~Rome~Bagabaldo, and Alexandre~M.~Bayen,
\newblock "The lord of the ring road: A review and evaluation of autonomous control policies for traffic in a ring road,"
\newblock \emph{ACM Trans. Cyber-Phys. Syst.}, vol. 6, no. 1, Jan 2022.

\bibitem{delle2019feedback}
Maria~Laura~Delle~Monache, Thibault~Liard, Anaïs~Rat, Raphael~Stern, Rahul~Bhadani, Benjamin~Seibold, Jonathan~Sprinkle, Daniel~B.~Work, and Benedetto~Piccoli,
\newblock "Feedback control algorithms for the dissipation of traffic waves with autonomous vehicles,"
\newblock In \emph{Computational Intelligence and Optimization Methods for Control Engineering}, pp. 275--299, 2019.

\bibitem{delle2019autonomous}
Maria~Laura~Delle~Monache, Jonathan~Sprinkle, Ramanarayan~Vasudevan, and Daniel~Work,
\newblock "Autonomous vehicles: From vehicular control to traffic control,"
\newblock In \emph{2019 IEEE 58th Conference on Decision and Control (CDC)}, pp. 4680--4696. IEEE, 2019.

\bibitem{CirclesNewsArticle}
Z.~Fu, K.~Manke, and A.~M.~Bayen,
\newblock "Massive CAV Experiment in Nashville Pits Machine Learning against Traffic Jams,"
\newblock 2023. Available: \url{https://pubsonline.informs.org/do/10.1287/orms.2023.02.05/full/}

\bibitem{Galvin2017}
R.~Galvin,
\newblock "Energy consumption effects of speed and acceleration in electric vehicles: Laboratory case studies and implications for drivers and policymakers,"
\newblock \emph{Transportation Research Part D: Transport and Environment}, vol. 53, pp. 234--248, June 2017.

\bibitem{GLOUDEMANS2023104311}
Derek~Gloudemans, Yanbing~Wang, Junyi~Ji, Gergely~Zachár, William~Barbour, Eric~Hall, Meredith~Cebelak, Lee~Smith, and Daniel~B.~Work,
\newblock "I-24 motion: An instrument for freeway traffic science,"
\newblock \emph{Transportation Research Part C: Emerging Technologies}, vol. 155, p. 104311, 2023.

\bibitem{Gong2018}
X.~Gong, Y.~Guo, Y.~Feng, J.~Sun, and D.~Zhao,
\newblock "Evaluation of the energy efficiency in a mixed traffic with automated vehicles and human controlled vehicles,"
\newblock In \emph{2018 IEEE Intelligent Transportation Systems Conference (ITSC)}, pp. 1981--1986, 2018.

\bibitem{gunter2020commercially}
George~Gunter, Derek~Gloudemans, Raphael~E.~Stern, Sean~McQuade, Rahul~Bhadani, Matt~Bunting, Maria~Laura~Delle~Monache, Roman~Lysecky, Benjamin~Seibold, Jonathan~Sprinkle, et al.,
\newblock "Are commercially implemented adaptive cruise control systems string stable?"
\newblock \emph{IEEE Transactions on Intelligent Transportation Systems}, vol. 22, no. 11, pp. 6992--7003, 2020.

\bibitem{khoudari2023reducing}
N.~Khoudari, S.~Almatrudi, R.~Ramadan, J.~Carpio, M.~Yao, K.~Butts, A.~M.~Bayen, J.~W.~Lee, and B.~Seibold,
\newblock "Reducing detailed vehicle energy dynamics to physics-like models,"
\newblock \emph{arXiv preprint arXiv:2310.06297}, 2023.

\bibitem{autonomie}
Argonne National Laboratory,
\newblock "Autonomie,"
\newblock 2023. Available: \url{https://www.anl.gov/taps/autonomie-vehicle-system-simulation-tool}

\bibitem{lee2021integrated}
J.~W.~Lee, G.~Gunter, R.~Ramadan, S.~Almatrudi, P.~Arnold, J.~Aquino, W.~Barbour, R.~Bhadani, J.~Carpio, F.~C.~Chou, et al.,
\newblock "Integrated framework of vehicle dynamics, instabilities, energy models, and sparse flow smoothing controllers,"
\newblock In \emph{Proceedings of the Workshop on Data-Driven and Intelligent Cyber-Physical Systems}, pp. 41--47. IEEE, 2021.

\bibitem{lee2024traffic_smoothing}
J.~W.~Lee, H.~Wang, K.~Jang, A.~Hayat, M.~Bunting, A.~Alanqary, W.~Barbour, Z.~Fu, X.~Gong, G.~Gunter, S.~Hornstein, A.~R.~Kreidieh, N.~Lichtlé, M.~W.~Nice, W.~A.~Richardson, A.~Shah, E.~Vinitsky, F.~Wu, S.~Almatrudi, F.~Althukair, R.~Bhadani, J.~Carpio, R.~Chekroun, E.~Cheng, M.~T.~Chiri, F.~C.~Chou, R.~Delorenzo, M.~Gibson, D.~Gloudemans, A.~Gollakota, J.~Ji, A.~Keimer, N.~Khoudari, M.~Mahmood, M.~N.~Baig, H.~N.~Z.~Matin, S.~Mcquade, R.~Ramadan, D.~Urieli, X.~Wang, Y.~Wang, R.~Xu, M.~Yao, Y.~You, G.~Zachár, Y.~Zhao, M.~Ameli, S.~Bhaskaran, K.~Butts, M.~Gowda, C.~Janssen, J.~Lee, L.~Pedersen, R.~Wagner, Z.~Zhang, C.~Zhou, D.~B.~Work, B.~Seibold, J.~Sprinkle, B.~Piccoli, M.~L.~Delle~Monache, and A.~M.~Bayen,
\newblock "Traffic smoothing via connected \& automated vehicles: A modular, hierarchical control design deployed in a 100-CAV flow smoothing experiment,"
\newblock \emph{IEEE Control Systems Magazine}, 2024.

\bibitem{lee2024traffic_control}
J.~W.~Lee, H.~Wang, K.~Jang, A.~Hayat, M.~Bunting, A.~Alanqary, W.~Barbour, Z.~Fu, X.~Gong, G.~Gunter, S.~Hornstein, A.~R.~Kreidieh, N.~Lichtlé, M.~W.~Nice, W.~A.~Richardson, A.~Shah, E.~Vinitsky, F.~Wu, S.~Xiang, S.~Almatrudi, F.~Althukair, R.~Bhadani, J.~Carpio, R.~Chekroun, E.~Cheng, M.~T.~Chiri, F.~C.~Chou, R.~Delorenzo, M.~Gibson, D.~Gloudemans, A.~Gollakota, J.~Ji, A.~Keimer, N.~Khoudari, M.~Mahmood, M.~N.~Baig, H.~N.~Z.~Matin, S.~Mcquade, R.~Ramadan, D.~Urieli, X.~Wang, Y.~Wang, R.~Xu, M.~Yao, Y.~You, G.~Zachár, Y.~Zhao, M.~Ameli, S.~Bhaskaran, K.~Butts, M.~Gowda, C.~Janssen, J.~Lee, L.~Pedersen, R.~Wagner, Z.~Zhang, C.~Zhou, D.~B.~Work, B.~Seibold, J.~Sprinkle, B.~Piccoli, M.~L.~Delle~Monache, and A.~M.~Bayen,
\newblock "Traffic control via connected and automated vehicles: An open-road field experiment with 100 CAVs,"
\newblock \emph{arXiv preprint arXiv:2402.17043}, February 2024.

\bibitem{liang1999optimal}
Chi-Ying~Liang and Huei~Peng,
\newblock "Optimal adaptive cruise control with guaranteed string stability,"
\newblock \emph{Vehicle System Dynamics}, vol. 32, no. 4--5, pp. 313--330, 1999.

\bibitem{lichtle2022deploying}
Nathan~Lichtlé, Eugene~Vinitsky, Matthew~Nice, Benjamin~Seibold, Dan~Work, and Alexandre~M.~Bayen,
\newblock "Deploying traffic smoothing cruise controllers learned from trajectory data,"
\newblock In \emph{2022 International Conference on Robotics and Automation (ICRA)}, pp. 2884--2890, 2022.

\bibitem{milanes2013cooperative}
Vicente~Milanés, Steven~E.~Shladover, John~Spring, Christopher~Nowakowski, Hiroshi~Kawazoe, and Masahide~Nakamura,
\newblock "Cooperative adaptive cruise control in real traffic situations,"
\newblock \emph{IEEE Transactions on Intelligent Transportation Systems}, vol. 15, no. 1, pp. 296--305, 2013.

\bibitem{Chassis_dyno}
openPR~Worldwide~Public~Relation,
\newblock "Automotive chassis dynamometers market 2022: Top key players in industry are slated to grow rapidly in the coming years till 2030 - horiba, meidensha, avl list, mts, rototest,"
\newblock August 2022. Available: \url{https://www.openpr.com/news/2714428/automotivechassis-dynamometers-market-2022-top-key-players}


\bibitem{shang2021impacts}
Mingfeng~Shang and Raphael~E.~Stern,
\newblock "Impacts of commercially available adaptive cruise control vehicles on highway stability and throughput,"
\newblock \emph{Transportation Research Part C: Emerging Technologies}, vol. 122, p. 102897, 2021.

\bibitem{stern2018dissipation}
Raphael~E.~Stern, Shumo~Cui, Maria~Laura~Delle~Monache, Rahul~Bhadani, Matt~Bunting, Miles~Churchill, Nathaniel~Hamilton, Hannah~Pohlmann, Fangyu~Wu, Benedetto~Piccoli, et al.,
\newblock "Dissipation of stop-and-go waves via control of autonomous vehicles: Field experiments,"
\newblock \emph{Transportation Research Part C: Emerging Technologies}, vol. 89, pp. 205--221, 2018.

\bibitem{sugiyama2008traffic}
Yuki~Sugiyama, Minoru~Fukui, Macoto~Kikuchi, Katsuya~Hasebe, Akihiro~Nakayama, Katsuhiro~Nishinari, Shin-ichi~Tadaki, and Satoshi~Yukawa,
\newblock "Traffic jams without bottlenecks—experimental evidence for the physical mechanism of the formation of a jam,"
\newblock \emph{New Journal of Physics}, vol. 10, no. 3, p. 033001, 2008.

\bibitem{treiber2000congested}
Martin~Treiber, Ansgar~Hennecke, and Dirk~Helbing,
\newblock "Congested traffic states in empirical observations and microscopic simulations,"
\newblock \emph{Physical Review E}, vol. 62, no. 2, p. 1805, 2000.

\bibitem{CirclesSite}
UC~Berkeley,
\newblock "Circles: Using deep reinforcement learning and self-driving cars to improve traffic flow and reduce energy consumption,"
\newblock 2019. Accessed January 25, 2023. Available: \url{https://circles-consortium.github.io/}

\bibitem{DriveScheduleEPA}
United~States~Environmental~Protection~Agency~(US~EPA),
\newblock "Dynamometer Drive Schedules,"
\newblock 2022. Available: \url{https://www.epa.gov/vehicle-and-fuel-emissions-testing/dynamometer-drive-schedule}

\bibitem{ToyotaEngineBenchmark}
US~EPA, National~Vehicle~and~Fuel~Emissions~Laboratory, National~Center~for~Advanced~Technology,
\newblock "2018 Toyota 2.5L A25A-FKS engine tier 3 fuel – test data package. Version 2020-07,"
\newblock 2020. Available: \url{https://www.epa.gov/vehicle-and-fuel-emissions-testing/benchmarking-advanced-low-emission-light-duty-vehicle-technology}

\bibitem{vinitsky2020energy}
E.~Vinitsky, N.~Lichtle, A.~Kreidieh, J.~Lee, A.~Velu, S.~Almatrudi, J.~Carpio, and A.~Bayen,
\newblock "Energy optimization of traffic at scale using reinforcement learning,"
\newblock Technical report, Regents of The University of California, The, August 2020.

\bibitem{wu2024modifying}
F.~Wu, J.~Carpio, M.~Bunting, M.~Nice, D.~Work, J.~Sprinkle, J.~Lee, S.~Hornstein, and A.~Bayen,
\newblock "Modifying adaptive cruise control systems for string stable stop-and-go wave control,"
\newblock \emph{IEEE Robotics and Automation Letters}, August 2024.

\bibitem{wu2019tracking}
Fangyu~Wu, Raphael~E.~Stern, Shumo~Cui, Maria~Laura~Delle~Monache, Rahul~Bhadani, Matt~Bunting, Miles~Churchill, Nathaniel~Hamilton, Benedetto~Piccoli, Benjamin~Seibold, et al.,
\newblock "Tracking vehicle trajectories and fuel rates in phantom traffic jams: Methodology and data,"
\newblock \emph{Transportation Research Part C: Emerging Technologies}, vol. 99, pp. 82--109, 2019.

\bibitem{xiao2010comprehensive}
Lingyun~Xiao and Feng~Gao,
\newblock "A comprehensive review of the development of adaptive cruise control systems,"
\newblock \emph{Vehicle System Dynamics}, vol. 48, no. 10, pp. 1167--1192, 2010.


\end{thebibliography}

\end{document}